\documentclass[11pt]{article}

\usepackage{ams}
\usepackage{amssymb}
\usepackage{graphicx}
\usepackage{caption}
\usepackage{pgf}
\usepackage{tikz}

\usepackage{amsthm}

\usetikzlibrary{arrows,automata}

\usepackage{latexsym}
\usepackage{amsfonts}
\usepackage[english]{babel}
\usepackage{xy}
\usepackage{fancyhdr}
\usepackage[a4paper,top=3cm,bottom=3.5cm,left=2.3cm,right=2.3cm,bindingoffset=5mm]{geometry}

\usepackage{setspace}

\newtheorem{lemma}{Lemma}
\newtheorem{corollary}{Corollary}
\newtheorem{proposition}{Proposition}

\newcommand{\DN}{\mbox{\boldmath $ D $}_N}
\newcommand{\sample}{\mbox{\boldmath $ s $}}

\newcommand{\YN}{\mbox{\boldmath $ Y $}_N}
\newcommand{\yinfty}{\mbox{\boldmath $ y $}_{\infty}}
\newcommand{\xinfty}{\mbox{\boldmath $ x $}_{\infty}}

\newcommand{\XN}{\mbox{\boldmath $ X $}_N}
\newcommand{\yN}{\mbox{\boldmath $ y $}_N}
\newcommand{\xN}{\mbox{\boldmath $ x $}_N}

\newcommand{\s}{\mbox{\boldmath $ s $}}

\def\ipot#1#2{{ \left \{ \matrix{ #1\cr #2\cr } \right. }}

\def\ipotc#1#2#3#4{{ \left \{ \matrix { #1\cr #2\cr #3\cr #4\cr } \right. }}
\definecolor{crimson}{rgb}{0.86, 0.08, 0.24}
\definecolor{darkraspberry}{rgb}{0.53, 0.15, 0.34}

\begin{document}

$ \, $

\bigskip

$ \, $

\centerline{\Large \noindent {\bf A unified principled framework for resampling}}

\centerline{\Large \noindent {\bf based on pseudo-populations: asymptotic theory}}

\bigskip

\centerline{Pier Luigi Conti\footnote{{Pier
Luigi Conti. Dipartimento di Scienze
Statistiche; Sapienza Universit\`{a} di Roma; P.le A.
Moro, 5; 00185 Roma; Italy. E-mail
pierluigi.conti@uniroma1.it}}}

\centerline{Daniela Marella\footnote{{Daniela
Marella. Dipartimento di Scienze della Formazione; Universit\`{a} Roma Tre;
Via D. Manin, 53; 00185 Roma; Italy. E-mail
daniela.marella@uniroma3.it}}}

\centerline{Fulvia Mecatti\footnote{{Fulvia
Mecatti. Dipartimento di Sociologia e Ricerca Sociale; Universit\`{a} di Milano-Bicocca;
Via Bicocca degli Arcimboldi, 8;
20126 Milano; Italy. E-mail
fulvia.mecatti@unimib.it}}}

\centerline{Federico Andreis\footnote{{Federico
Andreis. Department of Policy Analysis and Public Management; Universit\`{a}
Bocconi; Via Roentgen,1; 20136 Milano; Italy. E-mail
federico.andreis@unibocconi.it}}}

%\centerline{Affiliation}

%\centerline{Affiliation}

%\centerline{Address}

\bigskip

\centerline{\noindent {\bf Abstract}}

In this paper, a class of resampling techniques for finite populations  under complex sampling design is introduced. The basic idea on which they rest is a two-step
procedure consisting in: $(i)$ constructing a ``pseudo-population'' on the basis of sample data; $(ii)$ drawing
a sample from the predicted population according to an appropriate resampling design. From a logical point of view, this approach
is essentially based on the {\em plug-in} principle by Efron, at the ``sampling design level''. Theoretical justifications based on large sample theory
are provided. New approaches to construct pseudo populations based on various forms of calibrations are proposed. Finally, a simulation study is performed.

\bigskip


\bigskip

\newpage

\section{Introduction}
\label{sec:introduction}

The use of resampling methods in survey sampling has a long history, and several different techniques have been proposed in the literature.
The common starting point consists in observing that the classical bootstrap method, as proposed by \cite{efron:79}, does not work in survey sampling, because of the dependence among units due to the sampling design itself.
%The same comes true in other areas, for instance in time series, where variants of the bootstrap based on blocks of observations are used in order to compensate the dependence among observations (cfr. \cite{lahiri:03} and references therein).

Adaptations taking into account the non {\em i.i.d.} nature of the data are required when the sample is collected through a
%variable probability
 general sampling design, possibly assigning different  probability  to  every population unit to be included in the sample.
The literature on resampling from finite populations is mainly devoted to estimate variances of estimators; crf.  \cite{mashreg16}.
The main approaches are
essentially two:  {\em ad hoc} approaches and {\em plug in} approaches (cfr.  \cite{ranalli:12}, \cite{chauvet07} and references therein).

The basic idea of {\em ad hoc} approaches consists in resampling from the original sample through a special design, that accounts for the dependence among units. This approach is pursued in \cite{mccarthy85}, \cite{raowu88}, where the re-sampled data produced by the  ``usual''  $i.i.d.$ bootstrap are properly rescaled, as well as in \cite{sitter92}, \cite{beaumontpatak12}, \cite{chatterjee11}, \cite{contimar13}, where a ``rescaled bootstrap process'' based on asymptotic results is proposed. Among the {\em ad hoc} approaches we also quote the recent paper by \cite{antal:11}, where an ingenious mixed resampling design is proposed to account for the dependence among observations.

{\em Plug-in} approaches are based on the idea of ``expanding'' the sample to a ``pseudo-population'' that plays the role of a ``surrogate'' (actually an estimate) of the original one. Then, bootstrap samples are drawn from such a pseudo-population according to some appropriate resampling design: cfr. \cite{gross:80}, \cite{chao:85}, \cite{boothbutall94}, \cite{holmberg:98}, \cite{chauvet07}, as well as  \cite{mashreg16}.

Virtually all resampling techniques proposed for finite populations rest on the same justification: in case of linear statistics, the variance of the resampled statistic should match (or should be very close to) the ``usual'' variance estimator, possibly with approximated forms of the second order inclusion probabilities; cfr. \cite{antal:11}. This is far from the arguments commonly used to justify the classical bootstrap and its variants, that are based on asymptotic considerations involving the whole sampling distribution of a statistic (cfr., for instance, \cite{bickelfried81} and \cite{lahiri:03}): the asymptotic distribution of a bootstrapped
statistic should coincide with that of the ``original'' statistic. This argument is actually used in \cite{contimar13}.

In the present paper a class of resampling techniques for finite populations is proposed. It is based on a two-phase procedure. In the first phase, a
pseudo-population, that can be viewed as a
prediction of the population, is
constructed. In the second phase, a (re)sample is drawn from the pseudo-population.
In a sense, this approach parallels the {\em plug-in principle} by \cite{efron03}. The pseudo-population is  plugged in the sampling process,
and acts as a surrogate of the actual finite population. In other terms, the predicted population mimics the real population, and the (re)sampling
process from the predicted population mimics the
%actual
(original) sampling process from the real population.
From a formal point of view, the main justification of the whole procedure is based on large sample arguments.
In this sense, the approach pursued in the present paper  offers a principled framework for resampling from finite populations that parallels
 the arguments used for classical Efron's bootstrap of {\em i.i.d.} data.
For this reason, some preliminary
developments of large sample theory for finite populations are needed. In particular, we consider here high entropy sampling designs, similar to those studied in \cite{conti12}, \cite{contimar13}, but with an important addition: the possible relationships between the variable of interest and the design variables are explicitly taken into account. This dramatically changes the asymptotic results in  \cite{conti12}. As a matter of fact, the resampling
method defined in \cite{contimar13}, based on rescaling Efron's bootstrap, does not work when there is dependence between the variable of interest and the design variables.

The paper is organized as follows. Section \ref{sec:assumptions_prel}
 contains the basic assumptions on which the paper rests. Sections \ref{sec:hajek_asympt}, \ref{sec:hajek_asympt_stat_functionals} are devoted to asymptotic results for a wide class of estimators of appropriate population parameters.  Section \ref{sec:resampling_1} describes the proposed {\em predictive} resampling
 % resampling scheme
and the basic theoretical results.
In Section $\ref{sec:calib_1}$
%$ \ref{sec:htd_pseudopop}$
different %methods
 strategies  to construct pseudo-populations are introduced. In Section $\ref{sec:simul}$, such methods are compared {\em via} a Monte Carlo simulation study.  Conclusions are provided in Section $\ref{sec:conclus}$.
Technical lemmas and proofs are gathered in Appendix.

\section{Assumptions and preliminaries}
\label{sec:assumptions_prel}

Let ${\mathcal U}_N$ be a finite population of size $N$. A sample $\s$ is a subset of ${\mathcal U}_N$. For each unit $i \in {\mathcal U}_N$, let $D_i$ be a Bernoulli random variable (r.v.), such that $i$ is  (is not) in the sample $\s$ whenever  $D_i =1$ ($D_i =0$), so that $\s = \{ i \in {\mathcal U}_N : \; D_i =1 \}$.
Denote further by $\DN$  the $N$-dimensional r.v. of components $(D_1,\dots,D_N)$.
A (unordered, without replacement) sampling design $P$ is the probability distribution of the random vector $\DN$. The expectations  $\pi_i = E_P [ D_i ]$ and
$\pi_{ij} = E_P [ D_i  \, D_j ]$ are the first and second order inclusion probabilities, respectively. The suffix $P$ denotes the sampling design used to select the sample $\s$. The sample  size is  $n_s = D_1 + \cdots + D_N$.
In the sequel we will only consider fixed size sampling designs, such that $n_s \equiv  n$.

The first order inclusion probabilities are frequently chosen to be proportional to an auxiliary variable ${\mathcal X}$. In symbols:
 $\pi_i \propto x_i$,
where $x_i$ is the value of ${\mathcal X}$ for unit $i$
($i=1, \, \dots , \, N$). The rationale of this choice
is simple: if the values of the variable of interest are positively correlated with (or, even better, approximately proportional to) the values of the auxiliary variable, then the Horvitz-Thompson estimator of the population mean will be highly efficient.

For each unit $i$, let $p_i$ be a positive number, with  $p_1 + \cdots + p_N =n$. The {\em Poisson sampling design} ($Po$, for short) with parameters $p_1$, $\dots $, $p_N$ is characterized by the independence of the r.v.s $D_i$s, with $Pr_{Po} ( D_i =1 ) = p_i$. In symbols
\begin{eqnarray}
Pr_{Po} ( \DN ) = \prod_{i=1}^{N} p_i^{D_{i}} (1- p_i )^{1- D_{i}} .
\nonumber
\end{eqnarray}

The {\em rejective sampling}, or {\em normalized conditional Poisson sampling}
(cfr. \cite{hajek64}, \cite{tille06}) is obtained from the Poisson sampling by conditioning w.r.t. $n_s =n$. Using the suffix $R$ to denote
the rejective sampling design,
$E_{R} [D_i \, \vert n_s =n]$ is {\em not} generally equal to $p_i$, although they are asymptotically equivalent, as $N$ and $n$ increase (\cite{hajek64}). In \cite{chendempsterliu94} an algorithm is proposed to compute $p_i$s in terms of $\pi_i$s for the conditional Poisson sampling.

The rejective sampling design is characterized by a fundamental property: it possesses maximum entropy among all sampling designs of fixed size and
fixed first order inclusion probabilities (as shown in \cite{hajek81}), where the entropy of a sampling design $P$ is
\begin{eqnarray}
H (P) = E_P \left [ \log Pr_P ( \DN )  \right ] = \sum_{D_1 , \, \dots , \, D_N} Pr_P ( \DN ) \, \log \left (  Pr_P ( \DN ) \right ) .
\nonumber
\end{eqnarray}

The {\em Hellinger distance} between a sampling design $P$ and the rejective design is defined as
\begin{eqnarray}
d_H (P, \, P_R ) = \sum_{D_1 , \, \dots , \, D_N} \left (
\sqrt{Pr_P ( \DN )}  - \sqrt{Pr_{P_R} ( \DN )} \right )^2 .
\label{eq:hellinger}
\end{eqnarray}

From now on, the character of interest is denoted by $\mathcal{Y}$, and its value for unit $i$ by $y_i$.
$\mathcal{T}_1 , \, \dots , \, \mathcal{T}_L$ are the design variables, and
$t_{i1} , \, \dots , \, t_{iL}$ are their values for unit $i$. The design variables may include strata indicators,
as well as variables measuring cluster and unit characteristics (cfr. \cite{pfeffermann93}).
They are used to construct the sampling design, and to compute the sampling weights, {\em i.e.} the reciprocals of the
first order inclusion probabilities.

The basic assumptions on which the present paper relies are listed below.

\begin{itemize}
\item[A1.] $( {\mathcal U}_N ; \; N \geq 1)$ is a sequence of finite populations of increasing size $N$.
\item[A2.] For each $N$, $( y_i , \, t_{i1} , \, \dots , \, t_{iL} )$, $i=1, \, \dots , \, N$ are
 realizations of a superpopulation model $\{ ( Y_i , \, T_{i1} , \, \dots , \, T_{iL} ) , \; i=1 , \, \dots , \, N\}$ composed by
{\em i.i.d.} $(L+1)$-dimensional r.v.s.
The symbol $\mathbb{P}$ denotes the (superpopulation) probability distribution of r.v.s $( Y_i , \, T_{i1} , \, \dots , \, T_{iL} )$s, and
 $\mathbb{E}$, $\mathbb{V}$ are the corresponding mean and variance, respectively.
\item[A3.] For each population ${\mathcal U}_N$, sample units are selected according to a sample design  with positive first order inclusion probabilities
$\pi_1 $, $\dots $, $\pi_N$, and fixed sample size $n = \pi_1 + \cdots + \pi_N$. The first order inclusion probabilities are taken proportional to $x_i =
h ( t_{i1} , \, \dots , \, t_{iL} )$,  $h( \cdot )$ being an arbitrary positive function. To avoid complications in the notation, we will assume that
$\pi_i = n x_i / \sum_{i=1}^{N} x_i$ for each unit $i$.

Although the sample size $n$, the inclusion probabilities $\pi_i$s, and the r.v.s $D_i$s, as well, depend on $N$, in order to use a simple  notation  the symbols $n$, $\pi_i$, $D_i$ are used, instead of the more complete $n_N$, $\pi_{i,N}$, $D_{i,N}$. It is also assumed that
\begin{eqnarray}
\lim_{N, n \rightarrow \infty} \mathbb{E} [ \pi_i ( 1- \pi_i ) ] = d >0 .  \label{eq:quant_d}
\end{eqnarray}
%\noindent Note that $d \leq 1/4$ because $0 \leq \pi_i \leq 1$.
\item[A4.]
The sample size $n$ increases as the population size $N$ does, with
\begin{eqnarray}
\lim_{N \rightarrow \infty} \frac{n}{N} = f, \;\; 0 < f < 1 .
\nonumber
\end{eqnarray}
\item[A5.] For each population $( {\mathcal U}_N ; \; N \geq 1)$, let $P_R$ be
the rejective sampling design with inclusion probabilities $\pi_1$, $\dots$, $\pi_N$, and let $P$ be the actual sampling design
(with the same inclusion probabilities). Then
\begin{eqnarray}
d_H (P, \, P_R ) \rightarrow 0 \;\; {\mathrm as} \; N \rightarrow \infty , \;\; a.s.-{\mathbb{P}}. \nonumber
\end{eqnarray}
\item[A6.] $\mathbb{E} [ X_1^2 ] < \infty$, so that the quantity in $(  \ref{eq:quant_d} )$ is equal to:
\begin{eqnarray}
 d = f \left ( 1- \frac{\mathbb{E} [ X_1^2 ]}{\mathbb{E} [ X_1 ]^2} \right ) +
f (1-f) \frac{\mathbb{E} [ X_1^2 ]}{\mathbb{E} [ X_1 ]^2} > 0 .
\nonumber
\end{eqnarray}
\end{itemize}

Assumptions A2, A3 allow one to take into account  the possible dependence between the design variables and the study variable. Of course, this is
a key motivation for using non-simple, probability-proportional-to-size designs (dubbed  $\pi$ps sampling designs), where the dependence between $X_i$s and $Y_i$s is
important for the efficiency of the  estimation of the population mean (and other population parameters, as well).
Notice that assumptions A2, A3 do not %specify
 limit  the kind of dependence between $X_i$s and $Y_i$s, that can be completely general.

An obvious example of sampling designs satisfying A3 are $\pi$ps sampling designs, where the first order inclusion probability of unit $i$
is proportional to the value of a size measure.
Another elementary example is the stratified design. Assume that the population is subdivided into $L$ strata, composed by $N_1$, $\dots$, $N_L$
units, respectively
($N_1 + \cdots + N_L =N$).
Let further $w_l = N_l /N$, and let $p_1$, $\dots$, $p_L$ be arbitrary positive numbers such that $p_1 + \cdots + p_L =1$.
The stratified design drawing (by simple random sampling) $n_l = n p_l$ units from stratum $l$ ($= 1, \, \dots , \, L)$
can be considered as a special $\pi$ps sampling design where
the first order inclusion probability for unit $i$ is taken proportional to an auxiliary variable
(acting as a size measure) $x_i$ defined as
\begin{eqnarray}
x_i = \frac{p_{l}}{w_{l}} \;\; {\mathrm{if \; unit}} \; i \; {\mathrm{is \; within \; stratum}} \; l.
\label{eq:example_strata}
\end{eqnarray}
In fact, from $( \ref{eq:example_strata} )$ it easily follows that
\begin{eqnarray}
\pi_i = \frac{n p_{l}}{n} = \frac{n_{l}}{n} \;\; {\mathrm{if \; unit}} \; i \; {\mathrm{is \; within \; stratum}} \; l.
\end{eqnarray}
In particular, if $p_l = w_l$, then the sampling design reduces to stratified proportional sampling.

As discussed in \cite{conti12}, assumption A5 implies that the Kullback-Leibler divergence of the actual sampling design $P$ w.r.t. the rejective design
\begin{eqnarray}
\Delta_{KL} ( P \| P_R ) = H( P_R ) - H( P)   \label{eq:diff_entrop}
\end{eqnarray}
\noindent tends to zero as both $n$, $N$ increase. Hence, the sampling designs satisfying assumption A5 are essentially ``high entropy'' designs.
The importance of the high entropy property
of sampling designs  is discussed in  \cite{brewerdon03}, \cite{grafstrom10} and references therein.
Examples of sampling designs satisfying A5, as shown in \cite{berger98} and \cite{berger11}, are
simple random sampling,  successive sampling,  Rao-Sampford design, Chao  design, stratified design, two-stage design.

%{\color{purple} se c\'e la possibilit\'a che JNK Rao serva come referee, suggerisco di chiamarlo Rao-Sampford design.  In effetti Rao lo propose per $n=2$  e successivamente fu pubblicato  Sampford per $n$ qualunque.  Rao, J.N.K. (1965). On two simple schemes of unequal probability sampling without replacement. Journal of the Indian Statistical Association, 3, 173-180.}
% etc..

The {\em population distribution function} (p.d.f., for short) is:
\begin{eqnarray}
F_N (y) = \frac{1}{N} \sum_{i=1}^{N} I_{( y_{i} \leq y)} , \;\; y \in \mathbb{R}
\label{eq:popul_df}
\end{eqnarray}
\noindent where the indicator function $I_{( y_i \leq y )}$ is  equal to $1$ if $y_i \leq y$, and is equal to $0$ otherwise.

A {\em finite population parameter} is a functional (not necessarily real-valued) of the p.d.f.:
\begin{eqnarray}
\theta_N = \theta ( F_N ) .
\label{eq:popul_param}
\end{eqnarray}

The simplest (and widely used, as well) approach to estimate a finite population parameter of the form $( \ref{eq:popul_param} )$ consists in estimating first the p.d.f. ($\ref{eq:popul_df}$), and then in replacing
$F_N$ in  $( \ref{eq:popul_param} )$ by such an estimate. As an estimator of the p.d.f. $( \ref{eq:popul_df} )$ we consider here the
H\'{a}jek estimator:
\begin{eqnarray}
\widehat{F}_{H} (y) & = & \frac{\sum_{i=1}^{N} \frac{1}{\pi_{i}} D_i I_{(y_i \leq y)}}{\sum_{i=1}^{N} \frac{1}{\pi_{i}} D_i}
\label{eq:dfhajek}
\end{eqnarray}
\noindent which is a proper distribution function. It can be considered as the ``finite population version'' of the empirical distribution function, that plays a fundamental role in nonparametric statistics.
The finite population parameter $( \ref{eq:popul_param} )$ is then estimated by
\begin{eqnarray}
\widehat{\theta}_H = \theta \left ( \widehat{F}_{H} \right ) .
\label{eq:hajek_param}
\end{eqnarray}
In a sense, $( \ref{eq:hajek_param} )$ is the ``finite population version'' of {\em statistical functionals}.

The main task of Sections \ref{sec:hajek_asympt}, \ref{sec:hajek_asympt_stat_functionals} is to study the asymptotic properties of $( \ref{eq:dfhajek} )$,  $( \ref{eq:hajek_param} )$, respectively. In the sequel, the joint superpopulation d.f. of $( Y_i , \, X_i )$  will be denoted by
\begin{eqnarray}
H( y, \, x) = \mathbb{P} ( Y_i \leq y , \, X_i \leq x )
\label{eq:joint_sdf}
\end{eqnarray}
\noindent and the marginal superpopulation d.f.s of $Y_i$ and $X_i$ by
\begin{eqnarray}
F (y) = \mathbb{P} ( Y_i \leq y ) = H( y , \, + \infty ) , \;\;
G (x) = \mathbb{P} ( X_i \leq x ) = H( + \infty , \, x ) ,
\label{eq:marginal_sdf}
\end{eqnarray}
\noindent respectively. Furthermore, the notation
\begin{eqnarray}
K_{\alpha} (y) = \mathbb{E} \left [ \left . X_1^{\alpha} \, \right \vert Y_1 \leq y \right ] , \;\;
y \in {\mathbb{R}}, \; \alpha = 0, \, \pm 1, \, \pm 2 \label{eq:def_K}
\end{eqnarray}
\noindent will be used. Note that $K_{\alpha} ( + \infty ) = {\mathbb{E}} [ X_1^{\alpha} ]$.

\section{Estimating population distribution function}
\label{sec:hajek_asympt}

The goal of the present section is to derive the limiting distribution of the H\'{a}jek estimator $( \ref{eq:dfhajek})$, as the sample size and the population size increase.
To this purpose, consider the stochastic process $W^H_N = ( W^H_N (y) ; \; y \in \mathbb{R} )$, where
\begin{eqnarray}
W^H_N (y) = \sqrt{n} ( \widehat{F}_{H} (y) - F_N (y) ) ; \; y \in \mathbb{R} .
\label{eq:hajek_process}
\end{eqnarray}

It can be viewed as the finite population sampling version of the well-known empirical process.
The main result of the present section is Proposition $\ref{asympt_hajek}$, that establishes the weak convergence of $W^H_N$ to a Gaussian limiting process.
Proposition $ \ref{asympt_hajek}$  is in spirit similar to the main result in \cite{conti12}, but with fundamental differences that will be stressed in the sequel.

Before stating Proposition $ï¿½ \ref{asympt_hajek}$, we stress that in
our asymptotic approach the actual population $y_i$s and $x_i$s values are considered as {\em fixed}. The only source of variability is the sampling design, namely $\DN$.
If we let the population size $N$ go to infinity, we must also consider corresponding sequences $\yinfty = ( y_1, \, y_2 , \dots )$,
$\xinfty = ( x_1, \, x_2 , \dots )$ of
$y_i$s and $x_i$s values. The actual $\yN = ( y_1 , \, \dots , \, y_N)$, $\xN = ( x_1 , \, \dots , \, x_N)$ are the segments of the first $N$ $y_i$s,
$x_i$s in the sequences $\yinfty$, $\xinfty$, respectively. As $N$ increases,
$\yN$ tends to $\yinfty$ and $\xN$ tends to $\xinfty$.
By A2,
$\yinfty$, $\xinfty$ live in a probability space
$( (\mathbb{R}^2)^{\infty} , \, \mathcal{B} ( \mathbb{R}^2 )^{\infty} , \, \mathbb{P}^{\infty} )$, where $\mathcal{B} ( \mathbb{R}^2 )^{\infty}$ is the product Borel $\sigma$-field over $(\mathbb{R}^2)^{\infty}$, and
$\mathbb{P}^{\infty}$ is  the product measure  on $( \mathbb{R}^{\infty} , \, \mathcal{B} ( \mathbb{R} )^{\infty} )$ generated by $\mathbb{P}$.
The probability statements we consider  are of the form $Pr_P ( \cdot \vert \yN , \, \xN )$, with $N$ going to infinity. Conditioning w.r.t. $\yN$, $\xN$ means that
$y_i$s and $x_i$s are considered as fixed (although produced by a superpopulation model). The suffix $P$ means that the probability refers to the sampling design.
The results we will obtain hold for ``almost all'' sequences $\yinfty$, $\xinfty$ that the superpopulation model in A2 can produce, {\em i.e.}
for a set of sequences having $ \mathbb{P}^{\infty}$-probability 1. With a slight lack of precision, but more simply and intuitively, in the sequel
we will use the expression ``for almost all $y_i$s, $x_i$s values''.

\begin{proposition}
\label{asympt_hajek}
If the sampling design $P$ satisfies  assumptions A1-A6, with $\mathbb{P}$-probability $1$, conditionally on $\yN$, $\xN$
the sequence $ ( W_{N}^{H}  ; \; N \geq 1) $, converges weakly, in $D[ - \infty , \: + \infty ]$ equipped with the Skorokhod topology,
to a Gaussian process
$W^H  = (W^H (y) ; \; y \in \mathbb{R} )$ with zero mean function and covariance kernel
\begin{eqnarray}
C^H (y, \, t) & = &
 f \left \{ \frac{\mathbb{E} [ X_1 ]}{f} K_{-1} (y \wedge t) -1  \right \} F( y \wedge t)
- \frac{f^3}{d} \left ( 1 - \frac{K_1 (y)}{\mathbb{E} [ X_1 ]} \right )
\left ( 1 - \frac{K_1 (t)}{\mathbb{E} [ X_1 ]} \right ) F(y) F(t) \nonumber \\
\, & \, & - f
\left \{ \frac{\mathbb{E} [ X_1 ]}{f} \left ( K_{-1} (y) + K_{-1} (t) - \mathbb{E} \left [
X_1^{-1} \right ] - 1
\right )
\right \} F(y) F(t) ,
\label{eq:cov_ker_hajek}
\end{eqnarray}
\noindent with $d$ given by $( \ref{eq:val_d} )$.
\end{proposition}

When $X_i$ and $Y_i$ are independent, the covariance kernel $( \ref{eq:cov_ker_hajek} )$ reduces to
\begin{eqnarray}
f (A-1 ) ( F( y \wedge t) - F(y) F(t)) \nonumber
\end{eqnarray}
\noindent where
\begin{eqnarray}
A = \frac{\mathbb{E} [ X_1 ]}{f} \mathbb{E} [ X_1^{-1} ]
\label{eq:defin_A}
\end{eqnarray}
\nonumber is, with $\mathbb{P}$-probability 1, the limit of
\begin{eqnarray}
\frac{1}{N} \sum_{i=1}^{N} \frac{1}{\pi_{i}} \nonumber
\end{eqnarray}
\noindent as $N$ goes to infinity. Taking into account that $u \wedge v - uv$ is the covariance kernel of a
Brownian bridge $B = ( B(t) ; \; 0 \leq t \leq 1)$ ({\em i.e.} a Wiener process tied down at 1), we have thus proved the following corollary of Proposition $\ref{asympt_hajek}$.

\begin{corollary}
\label{asympt_hajek_indip}
If the sampling design $P$ satisfies  assumptions A1-A6, and if $X_i$ and $Y_i$ are independent, with $\mathbb{P}$-probability $1$, conditionally on $\yN$, $\xN$
the sequence $ ( W_{N}^{H}  ; \; N \geq 1) $, converges weakly, in $D[ - \infty , \: + \infty ]$ equipped with the Skorokhod topology,
to a Gaussian process that can be represented in the form
\begin{eqnarray}
(f (A-1) B( F( y)) ; \; y \in \mathbb{R} )
\label{eq:limit_bridge}
\end{eqnarray}
\noindent  as $N$ goes to infinity, where $B$ is a Brownian bridge and $A$ is given by $( \ref{eq:defin_A} )$.
\end{corollary}

Corollary $\ref{asympt_hajek_indip} $ essentially coincides with Proposition 2 in \cite{conti12}.
Proposition $\ref{asympt_hajek}$ is new. Due to the choice of the inclusion probabilities in A3, {\em i.e.} $\pi_i \propto x_i$,
and due to the possible dependence between $X_i$ and $Y_i$ (that usually comes true in practice),
the limiting Gaussian process is {\em not} proportional to a Brownian bridge.
Proposition $\ref{asympt_hajek}$ shows how the dependence between variable of interest and
design variables affects the covariance kernel of the Gaussian limiting law of $W_N^H $. If compared to
Proposition 2 in \cite{conti12}, its main consequence is that, whenever there is some kind of
dependence between the design variables (or, equivalently, the sampling weights) and the variable of interest, the empirical process
$( \ref{eq:hajek_process} )$ does not converge weakly to a Brownian bridge, but to a Gaussian process with a covariance kernel
 having a complicate form, depending on the
relationships between the character of interest and the design variables. The form of such a relationship is usually unknown.

From the proof of Proposition $\ref{asympt_hajek}$ it is clear that the assumption of independence and identical distribution of r.v.s
$(Y_i , \, X_i)$ is far from being necessary. It can be replaced by forms of dependence that
admit the strong law of large numbers.

Before ending the present section we note, {\em in passim}, that Proposition $ \ref{asympt_hajek}$ implies that,  with $\mathbb{P}$-probability $1$, conditionally on $\yN$, $\xN$:
\begin{eqnarray}
\vert \widehat{F}_H (y)  - F_N (y) \vert \stackrel{p}{\rightarrow} 0 \;\; {\mathrm{as}} \; N \rightarrow \infty
\label{eq:gliv1}
\end{eqnarray}
\noindent where the symbol $\stackrel{p}{\rightarrow}$ denotes the convergence in probability w.r.t. the sampling design (or better, w.r.t. the sequence of sampling designs in A3).
Using the same arguments as the proof of the Glivenko-Cantelli theorem, it is not difficult to prove the following further result.
\begin{proposition}
\label{glivenko_hajek}
If the sampling design $P$ satisfies  assumptions A1-A6, with $\mathbb{P}$-probability $1$, conditionally on $\yN$, $\xN$,
$
\sup_y \vert \widehat{F}_H (y)  - F_N (y) \vert
$
 converges to $0$ in probability w.r.t. the sampling design.
\end{proposition}

\noindent {\bf Remark.} Propositions $\ref{asympt_hajek}$, $\ref{glivenko_hajek}$, also hold when the inclusion probabilities $\pi_i$s depend on $y_i$s, {\em i.e.} when the sampling design is {\em informative}. This is true, in particular, when, for units in the sample, $\pi_i$s only depend on $y_i$s
of sample units, {\em i.e.} for adaptive designs. Even if this would be a point of
separate interest, we do not pursue in this direction.

\section{Estimating finite population parameters}
\label{sec:hajek_asympt_stat_functionals}

The goal of the present section is to study the large sample distribution of estimators of the finite population parameters that are functions of
p.d.f. $F_N ( \cdot )$. In particular, we concentrate on estimators of the form $( \ref{eq:hajek_param} )$. In a sense, the results of the present section can be viewed as a finite population version of the theory of statistical functionals, that mainly refers to the case of {\em i.i.d.} observations (cfr. \cite{vandervaart98}, Ch. 20).

The appropriate tool to study asymptotic properties of statistical functionals is the notion of Hadamard-differentiability.
Let $\theta ( \cdot ): \: l^{\infty} [ - \infty , \, + \infty ] \rightarrow E$ be a map having as domain the normed space $l^{\infty} [ - \infty , \, + \infty ]$ (endowed with the sup-norm), and taking values on an appropriate normed space $E$ with norm $\| \cdot \|_E $. The map $\theta ( \cdot )$ is Hadamard-differentiable at $F$ if there exists a continuous linear mapping $\theta^{\prime}_F : \: l^{\infty} [ - \infty , \, + \infty ] \rightarrow E$ such that
\begin{eqnarray}
\left \| \frac{\theta (F + t h_{t} ) - \theta (F)}{t} - \theta^{\prime}_F (h) \right \|_E
\rightarrow 0 \;\; as \; t \downarrow 0 , \; for \; every \;
h_t \rightarrow h .
\label{eq:had_diff}
\end{eqnarray}
The quantity $\theta^{\prime}_F ( \cdot )$ is the {\em Hadamard derivative} of $\theta$ at $F$.
Let  us consider the (sequence of) stochastic process
\begin{eqnarray}
T^H_N = \sqrt{n} \left ( \theta ( \widehat{F}_H ) - \theta ( F_N ) \right ) , \;\; N \geq 1.
\label{eq:tndef}
\end{eqnarray}
\noindent
In view of Theorem 20.8 in \cite{vandervaart98} and Proposition $\ref{asympt_hajek}$, the following result holds.

\begin{proposition}
\label{asy_hadamard}
Suppose that  $\theta ( \cdot )$ is (continuously) Hadamard-differentiable at $F$, with Hadamard derivative $\theta^{\prime}_F ( \cdot )$.
Under  assumptions A1-A6, with $\mathbb{P}$-probability $1$, conditionally on $\yN$, $\xN$, the sequence
$(T^H_N  ; \; N \geq 1) $ converges weakly to $\theta^{\prime}_F ( W^H )$, as $N$ increases.
\end{proposition}

Proposition $\ref{asy_hadamard}$ essentially provides,  under mild conditions, an asymptotic approximation for the sampling distribution of
$T^H_N  $. In particular, if $\theta$ is real-valued, since $\theta^{\prime}_F ( \cdot )$ is linear and $W^H $ is a Gaussian process, the law of $\theta^{\prime}_F ( W^H )$ is normal with mean zero and variance
\begin{eqnarray}
\sigma^2_\theta = \mathbb{E} \left [  \theta^{\prime}_F ( W^H )^2 \right ] .
\label{eq:asympt_var_theta}
\end{eqnarray}

\section{A class of resampling procedure and its basic properties}
\label{sec:resampling_1}

%\subsection{General aspects}
%\label{sec:resampling_1a}

The goal of this section is to introduce a unified class of resampling procedures working under the sampling designs considered in Section $\ref{sec:assumptions_prel}$, and that provides an approximation
of the sampling distribution of estimators of the form $( \ref{eq:hajek_param} )$.

The main theoretical justification we will provide is based on asymptotic arguments: the probability
distribution of the estimator $\theta ( \widehat{F}_H )$ and its approximation based on resampling both converge to the {\em same} limit. This is actually the main argument in favour of the
classical (nonparametric) bootstrap for {\em i.i.d.} data: cfr., for instance, \cite{bickelfried81}. The results of the present section can be viewed as
an attempt to reconciliate
%{\color{purple} a proposition (an effort?) to realign}
the arguments used in sampling finite populations with those used in classical nonparametric statistics.

The first attempt to define a resampling technique for finite populations based on asymptotic distribution theory is in
\cite{chatterjee11} for simple random sampling, and in
\cite{contimar13} for general designs. In the latter paper, a technique
based on rescaling classical bootstrap is proposed, and its properties are studied. However, two points have to be
stressed. The first one is that the technique developed in \cite{contimar13} is specifically designed to estimate quantiles. The second one is that it is fully justified from
an asymptotic point of view only when
there  are no relationships between $\pi_i$s (and hence $x_i$s) and $y_i$s. In other words, the rescaled bootstrap proposed in \cite{contimar13} does not work when the dependence
between $y_i$s and $x_i$s cannot be neglected.

In view of the above remarks, in this section
%we attempt at introducing
 a new resampling  algorithm
 %technique
 for finite population  is introduced, that works
\begin{itemize}
\item[$(i)$] for {\em general} estimators $\theta ( \widehat{F}_H )$ of general population parameters $\theta (F_{N})$;
\item[$(ii)$] when $x_i$s ({\em i.e.} the design variables) and $y_i$s ({\em i.e.} the variable of interest) are related by some kind of dependence.
No special assumption is made on the relationship between $x_i$s and $y_i$s, apart from its (possible) existence.
\end{itemize}

As already said in the introduction, the class of resampling techniques rests on a two-phase procedure. In the first phase,
on the basis of the sampling data a pseudo-population, {\em i.e.} a prediction of the  actual population is constructed.
In the second phase, a sample of size $n$ (the same as the ``original'' one) is drawn from  such a pseudo-population, according to a $\pi ps$ sample design $P^*$ (the {\em} resampling design) with inclusion probabilities appropriately chosen and satisfying the entropy condition A5.
The resampling design $P^*$ is not assumed to coincide with the sampling design $P$ used to collect data from the actual population.

From now on, the term {\em sampling design} $P$ denotes the sampling procedure drawing $n$ units from the ``original''
population $\mathcal{U}_N$. The {\em resampling design} $P^*$ is the sampling procedure drawing $n$ units from the predicted (pseudo-)population $\mathcal{U}^*_{N^*}$. Details of the two phases on which the resampling procedure relies are in Sections \ref{sec:resampling_1b},
\ref{sec:resampling_1c}.

\subsection{Phase 1: Pseudo-population}
\label{sec:resampling_1b}

A pseudo-population, {\em i.e.} a design-based population predictor of the population, is
\begin{eqnarray}
\left \{ ( N^{*}_i D_i , \, y_i , \, x_i  ) ; \; i  =1, \, \dots , \, N \right \}
\label{eq:predict_des}
\end{eqnarray}
\noindent where $N^{*}_i$s are integer-valued r.v.s, with (joint) probability distribution $P_{pred}$.
In practice, $( \ref{eq:predict_des} )$ means that
$N^{*}_i D_i$ population units are predicted to have $y$-value equal to $y_i$ and
$x$-value equal to $x_i$, for each sample unit $i$.
In the sequel, the familiar bootstrap symbols $y^{*}_k$, $x^{*}_k$ will be used to denote the $y$-value and $x$-value of unit $k$
of the predicted population, respectively. Of course $N^{*}_i$ units of the predicted population satisfy the relationships $y^{*}_k = y_i$,
$x^{*}_k = x_i$, $i \in \s$.
The d.f. of the pseudo-population is equal to
\begin{eqnarray}
F^{*}_{N^{*}} (y) = \frac{1}{N^{*}} \sum_{k=1}^{N^{*}} I_{(y^{*}_{k} \leq y)}
=
 \sum_{i=1}^{N} \frac{N^{*}_{i}}{N^{*}} D_i I_{(y_{i} \leq y)} , \;\; y \in {\mathbb{R}}
\label{eq:df_predicted}
\end{eqnarray}
\noindent where
\begin{eqnarray}
N^{*} = \sum_{i=1}^{N} N^{*}_i D_i .
\label{eq:defnstar}
\end{eqnarray}
\noindent is the total number of units of the pseudo-population.

As far as the terms $N^*_i$ are concerned, we will make the following assumptions on expectations, variances, covariances
w.r.t. $P_{pred}$.
\begin{itemize}
\item[P1.]
$E [ N^*_i \, \vert \DN , \, \YN, \, \XN ] = \pi_i^{-1} D_i  K_{1N} ( \DN , \, \YN , \, \XN )$
\item[P2.]
$V ( N^*_i \, \vert \DN , \, \YN, \, \XN ) \leq \pi_i^{-1} D_i  K_{2N} ( \DN , \, \YN , \, \XN ) $
\item[P3.]
$ \left \vert C ( N^*_i , \, N^*_h  \, \vert \DN , \, \YN, \, \XN ) \right \vert \leq \frac{c}{N}
  \pi_i^{-1}  \pi_h^{-1} D_i  D_h K_{3N} ( \DN , \, \YN , \, \XN )$ $i \neq h$
\end{itemize}
\noindent $c$ being an appropriate constant, with
\begin{eqnarray}
 K_{1N} ( \DN , \, \YN , \, \XN ) \rightarrow 1
\label{eq:cond_conv}
\end{eqnarray}
\noindent
and $ K_{jN} ( \DN , \, \YN , \, \XN )$, $j=2, \, 3$ are bounded in  probability, conditionally on
$ \DN , \, \YN , \, \XN$,
as $N$ increases. The symbol $\rightarrow$ in $( \ref{eq:cond_conv} )$ denotes convergence in probability w.r.t. $\DN$ and for
almost all $y_i$s, $x_i$s.

\subsection{Phase 2: Resampling design from the pseudo-population}
\label{sec:resampling_1c}

%The resampling procedure we consider is described below.

%{\sf{
%\begin{itemize}
%\item[1.] Generate a predicted population $( \ref{eq:predict_des} )$ of $N^{*}$ units. Denote by $y^{*}_k$, $x^{*}_k$ the %$y$-value and $x$-value of unit $k$ of the predicted population, respectively. Of course $N^{*}_i$ units of of the predicted %population satisfy the relationships $y^{*}_k = y_i$,
%$x^{*}_k = x_i$, $i \in \s$.
%\item[2.]
%Draw a sample $\s^*$ of size $n$ from the predicted population defined in phase 1, on the basis of a resampling design $P^*$ %with first order inclusion probabilities $\pi_k^* = n x^*_k / \sum_{h=1}^{N^{*}} x^*_h$ and satisfying assumption A5.
%\end{itemize}
%}}

In phase 2 a sample  $\s^*$ of size $n$ (the same as the original sample) is selected from the predicted population according to a resampling design $P^{*}$ with first order inclusion probabilities $\pi_k^* = n x^*_k / \sum_{h=1}^{N^{*}} x^*_h$ and satisfying the entropy assumption A5. The H\'{a}jek estimator of the d.f. of the predicted population
$F^{*}_{N^{*}} (y)$ is equal to
\begin{eqnarray}
\widehat{F}^*_H (y) = \frac{\sum_{k=1}^{N^{*}} \frac{D^{*}_{k}}{\pi^{*}_{k}} I_{( y^{*}_{k} \leq y)}}{\sum_{k=1}^{N^{*}} \frac{D^{*}_{k}}{\pi^{*}_{k}}} .
\label{eq:hajek_resampled}
\end{eqnarray}
where $D^{*}_{k} =1$ if the unit $k$ of the predicted population is drawn, and $D^{*}_k =0$ otherwise.

\begin{proposition}
\label{lemma_6a}
Under assumptions A1-A6, P1-P3, for almost all $y_i$s, $x_i$s values, and in probability w.r.t. $\DN$, \begin{eqnarray}
\frac{N^{*}}{N} \rightarrow 1 \;\; {\mathrm{in \; probability \; w.r.t. \;}}  P_{pred}
\label{eq:equiv_N*}
\end{eqnarray}
\noindent as $N$ goes to infinity.
\end{proposition}

The statement ``in probability w.r.t. $\DN$'' means that the set of $\DN$s values, for which Lemma $\ref{lemma_6}$ in Appendix holds,  possesses a
probability tending to 1 as $N$ increases.

%At this point, we are in a position to establish the main result of the present section.

Define now the ``resampled version'' of the processes
$W_{N}^{H} $ $( \ref{eq:hajek_process} )$ and $T^H_N $ $( \ref{eq:tndef} )$, namely
\begin{eqnarray}
W_{N}^{H*}  &= & \left ( \sqrt{n} ( \widehat{F}^{*}_H (y) - F^{*}_{N^{*}} (y) )  , \:\: y \in \mathbb{R} \right ) , \;\; N \geq 1  ; \label{eq:empirical_resampled_hajek}\\
T_{N}^{H*}  &=&\sqrt{n} ( \theta ( \widehat{F}^{*}_H ) - \theta ( F^{*}_{N^{*}}  )) ,   \;\; N \geq 1 . \label{eq:empirical_resampled_functional_hajek}
\end{eqnarray}

Proposition \ref{asympt_ricamp_hajek} contains the main result of the present section and it can be proved essentially with the same technique as Propositions $\ref{asympt_hajek} $, $\ref{asy_hadamard}$, respectively.

\begin{proposition}
\label{asympt_ricamp_hajek}
Suppose that the sampling design $P$ and the resampling design $P^*$ both satisfy  assumptions A1-A6, and that conditions
P1-P3 are fulfilled. Conditionally on $\yN$, $\xN$, $\DN$, $( D_1 N^*_1 , \, \dots , \, D_N N^*_N)$,
the following statements hold.
\begin{itemize}
\item[R$1$.]
The sequence $ ( W_{N}^{H*}  ; \; N \geq 1) $ converges weakly, in $D[ - \infty , \: + \infty ]$ equipped with the Skorokhod topology,
to a Gaussian process
$W^H  $ with zero mean function and covariance kernel $( \ref{eq:cov_ker_hajek} )$.
\item[R$2$.] If $\theta ( \cdot )$ is continuously Hadamard differentiable at $F$, then  $( T_{N}^{H*}  ; \; N \geq 1) $
converges weakly to $\theta^{\prime}_F ( W^H  )$, as $N$ increases.
\end{itemize}
In both $R1$, $R2$ weak convergence takes place for a set of $y_i$s, $x_i$s having $\mathbb{P}$-probability 1,
and for a set of $\DN$s and $( N^*_1 , \, \dots , \, N^*_N)$ of probability tending to $1$.
\end{proposition}

Proposition $\ref{asympt_ricamp_hajek}$  shows that the resampled process $W_{N}^{H*} $ ($T_{N}^{H*}$)  possesses the same limiting law as the ``original'' process
$W_{N}^{H} $ ($T_{N}^{H}  $) in Proposition $\ref{asympt_hajek} $ ($\ref{asy_hadamard}$). In other words, the proposed resampling procedure asymptotically recovers the probability law of
$W_{N}^{H} ( \cdot )$ and $T_{N}^{H} ( \cdot )$, respectively .

Proposition $\ref{asympt_ricamp_hajek}$ does not require that the resampling design coincides with the original sampling design, as in \cite{holmberg:98}. The essential required conditions are two: $(i)$
the predicted population is constructed as in phase 1; $(ii)$ the first order inclusion probabilities of the resampling design are proportional to the corresponding $x_i$ values, exactly as
the original sampling design. Intuitively speaking, this happens because both the original sampling design and the resampling design possess high entropy, and in this case their limiting behaviour essentially depends on the first order inclusion probabilities.

In Proposition  $\ref{asympt_ricamp_hajek}$ the probability distribution of $ W_{N}^{H*} $ ( $ T_{N}^{H*} $) is considered  conditionally on $\yN$, $\xN$, $\DN$, $( N^*_1 , \, \dots , \, N^*_N)$.
In other terms, the predicted population is considered as {\em fixed} (as well as $\yN$, $\xN$, $\DN$), and the only source of variability is the resampling design from the
predicted population.
Using Lemmas $1.1$, $1.2$ in \cite{csorgorosalsky03}, it is possible to see that the same result also holds when one considers the distribution of $ W_{N}^{H*} $ ( $ T_{N}^{H*} $)
conditionally on $\yN$, $\xN$, $\DN$. In this case only $\yN$, $\xN$, $\DN$ are considered as fixed, and there are {\em two} sources of variability: $(i)$
the variability of the process generating the predicted population  and $(ii)$ the variability of the resampling design from the predicted population. More precisely, the following proposition (that can be proved
with the same reasoning as in \cite{csorgorosalsky03}, based on Lemmas $1.1$, $1.2$ in the above paper) holds true.

\begin{proposition}
\label{asympt_ricamp_hajek_uncond}
Suppose the sampling design $P$ and the resampling design $P^*$ satisfy  assumptions A1-A6. Conditionally on $\yN$, $\xN$, $\DN$,
the following statements hold.
\begin{itemize}
\item[U$1$.]
The sequence $ ( W_{N}^{H*}  ; \; N \geq 1) $  converges weakly, in $D[ - \infty , \: + \infty ]$ equipped with the Skorokhod topology,
to a Gaussian process
$W^H  $ with zero mean function and covariance kernel $( \ref{eq:cov_ker_hajek} )$.
\item[U$2$.] If $\theta ( \cdot )$ is continuously Hadamard differentiable at $F$, then  $( T_{N}^{H*}  ; \; N \geq 1) $
converges weakly to $\theta^{\prime}_F ( W^H )$, as $N$ increases.
\end{itemize}
In both $U1$, $U2$ weak convergence takes place for a set of $y_i$s, $x_i$s having $\mathbb{P}$-probability 1,
and for a set of $\DN$s of probability tending to $1$.
\end{proposition}

The main consequence of  Propositions $\ref{asympt_ricamp_hajek}$, $\ref{asympt_ricamp_hajek_uncond}$ is that in generating the bootstrap samples two different approaches can be followed:
\begin{itemize}
\item [1.1] \emph{Conditional Approach}:  construct a predicted population and generate $M$  bootstrap samples $\s^*$ from it;
\item [1.2] \emph{Unconditional approach}: construct  $M$ predicted populations and generate one bootstrap sample $\s^*$ from each of them.
\end{itemize}
Clearly, the unconditional approach is  more computationally intensive and time consuming  than the conditional one.

The basic steps of the resampling procedure are described below. To simplify the notation, in the sequel we will assume that
$\theta ( \cdot )$ is real-valued, {\em i.e.} we will consider the case of scalar population parameters.

\begin{itemize}
\item[\emph{Step 1}] Generate $M$ independent bootstrap samples $\s^*$ of size $n$ on the basis of the two-phase procedure described above.
\item[\emph{Step 2}] For each bootstrap sample,  compute the corresponding H\'ajek estimator $( \ref{eq:hajek_resampled} )$. They will be denoted by $\widehat{F}^{*}_{H,m} (y)$, $m=1, \, \dots , \, M$.
\item[\emph{Step 3}]Compute the corresponding estimates of $\theta ( \cdot )$:
\begin{eqnarray}
\widehat{\theta}^{*}_{m}  = \theta ( \widehat{F}^{*}_{H,m} ) ; \;\; m=1, \, \dots , \, M.
\nonumber
\end{eqnarray}
\item[\emph{Step 4}] Compute the $M$ quantities
\begin{eqnarray}
Z^{*}_{n,m} = \sqrt{n} \left ( \widehat{\theta}^{*}_{m} - \theta ( F^*_{N^{*}} ) \right )
= \sqrt{n} \left ( \theta ( \widehat{F}^{*}_{H,m} ) - \theta ( F^*_{N^{*}} ) \right )
; \;\;
m=1, \, \dots , \, M.
\label{eq:def_z}
\end{eqnarray}
\item[\emph{Step 5}]
Compute the variance of $( \ref{eq:def_z} )$:
\begin{eqnarray}
\widehat{S}^{2*} = \frac{1}{M-1} \sum_{m=1}^{M} \left ( Z^{*}_{n,m} - \overline{Z}^{*}_M  \right )^2
= \frac{n}{M-1} \sum_{m=1}^{M} \left ( \widehat{\theta}^{*}_{m} - \overline{\theta}^{*}_M  \right )^2
\label{eq:resamp_var}
\end{eqnarray}
\noindent where
\begin{eqnarray}
\overline{Z}^{*}_M = \frac{1}{M} \sum_{m=1}^{M}  Z^{*}_{n,m} , \;\;
\overline{\theta}^{*}_M = \frac{1}{M} \sum_{m=1}^{M}   \widehat{\theta}^{*}_{m} .
\nonumber
\end{eqnarray}
\end{itemize}

Denote further by
\begin{eqnarray}
\widehat{R}^{*}_{n,M} (z) = \frac{1}{M} \sum_{m=1}^{M} I_{( Z^{*}_{n,m} \leq z )} , \;\; z \in \mathbb{R}
\label{eq:resampling_edf}
\end{eqnarray}
\noindent the empirical distribution function of $Z^{*}_{n,m}$s, and by
\begin{eqnarray}
\widehat{R}^{* -1}_{n,M} (p) = \inf \{ z : \; \widehat{R}^{*}_{n,M} (z) \geq p \} , \;\; 0 <p<1
\label{eq:resampling_quantiles}
\end{eqnarray}
\noindent the corresponding $p$th quantile.

The empirical d.f. $( \ref{eq:resampling_edf} )$ is essentially an approximation of the (resampling)
distribution of $T^{H*}_N $ as defined by equation (\ref{eq:empirical_resampled_functional_hajek}). In Proposition  $\ref{prop_resampling_comp} $
it is shown that it converges to the same limit as the d.f. of $T^{H*}_N $, and
that a similar result holds for the quantiles $( \ref{eq:resampling_quantiles} )$.

\begin{proposition}
\label{prop_resampling_comp}
Suppose that  assumptions A1-A6 are satisfied,  let
$\sigma^2_\theta $ be defined as in $( \ref{eq:asympt_var_theta} )$,
let $\Phi_{0, \sigma^{2}_\theta}$ be a normal distribution function with expectation $0$ and variance
$\sigma^2_\theta$, and let $\Phi_{0, \sigma^{2}_\theta}^{-1} (p)$ be the $p$-quantile of $\Phi_{0, \sigma^{2}_\theta}$
({\em i.e.} the unique solution of $\Phi_{0, \sigma^{2}_\theta} (z) = p$), $0 < p < 1$.

For almost all $y_i$s, $x_i$s values, and in probability w.r.t. $\DN$, $( N^*_1 , \, \dots , \, N^*_N )$,
conditionally on $\yN$, $\xN$, $\DN$, $( N^*_1 , \, \dots , \, N^*_N )$,
the following results hold:
\begin{eqnarray}
\, & \, & \sup_{z} \left \vert \widehat{R}^{*}_{n,M} (z) - \Phi_{0, \sigma^2_{\theta}} (z) \right \vert \stackrel{a.s.}{\rightarrow} 0 \,
; \label{eq:risult_1} \\
\, & \, & \widehat{R}^{*-1}_{n,M} (p) \stackrel{a.s.}{\rightarrow} \Phi_{0, \sigma^{2}_\theta}^{-1} (p) , \;\; \forall \,
0<p<1 \label{eq:risult_2}
\end{eqnarray}
\noindent as $M$, $N$ go to infinity.

In addition, if the sequence $\left ( Z^{*}_{m} - \overline{Z}^{*}_M  \right )^2 $ is dominated by a r.v. $U$  with finite expectation,  {\em i.e.}
$\left ( Z^{*}_{m} - \overline{Z}^{*}_M  \right )^2 \leq U$ for each $n$, $N$ and $M$, then
in probability w.r.t.  $\yN$, $\xN$, $\DN$, $( N^*_1 , \, \dots , \, N^*_N )$,
 conditionally on $\yN$, $\xN$, $\DN$, $( N^*_1 , \, \dots , \, N^*_N )$  it yields
\begin{eqnarray}
\widehat{S}^{2*} \rightarrow \sigma^2_\theta \;\; {\mathrm{as}} \; M, \, N  \rightarrow \infty
\label{eq:risult_3}
\end{eqnarray}
\noindent where convergence in $( \ref{eq:risult_3} )$ is in probability w.r.t. resampling
replications.
\end{proposition}

The main  consequences of Proposition $\ref{prop_resampling_comp}$ are two. First of all, {\em the estimator $\widehat{S}^{2*} $
is a consistent estimator of the variance of $\theta ( \widehat{F}_H )$}. In the second place, {\em the confidence intervals}
\begin{eqnarray}
\, & \, &
\left [ \widehat{\theta}_H - n^{-1/2} R^{* -1}_{n,M} (1- \alpha /2 ) , \;
\widehat{\theta}_H - n^{-1/2} R^{* -1}_{n,M} ( \alpha /2 )
\right ] \label{conf_int_percentile} \\
\, & \, &
\left [ \widehat{\theta}_H - n^{-1/2}  z_{\alpha /2} \widehat{S}^{*} , \;
\widehat{\theta}_H + n^{-1/2}  z_{\alpha /2} \widehat{S}^{*} \right ]
\label{eq:conf_int_norm}
\end{eqnarray}
\noindent {\em both possess asymptotic  confidence level $1- \alpha$} as $N$ and $M$ increase.

\section{ Some strategies for constructing pseudo-populations}
\label{sec:calib_1}

In view of Proposition $\ref{asympt_ricamp_hajek}$, all techniques to construct a pseudo-population are asymptotically equivalent,
provided that they satisfy conditions $P1$-$P3$ of Section $\ref{sec:resampling_1b}$. In this sense in  the present  paper  a unified approach for resampling  based on pseudo-populations is given.
However  in practical applications, {\em i.e.}  for finite $n$, a crucial aspect that would potentially affect the performance of resampling,
%based on pseudo-population,
 is how the pseudo-population is constructed.
The idea behind pseudo-populations is simple: as the sample and population sizes increase, the pseudo-population
tends to be ``similar'' to the real finite population. Hence, it would be intuitive to use a pseudo-population that is
as similar as possible to the actual population. In a sense, the pseudo-population should be somehow calibrated w.r.t.
the population.
Such an intuition can be put into practice in several ways. In the present section  some proposals
based on different calibration approaches are illustrated, which lead to different pseudo-populations.

\subsection{ Horvitz-Thompson pseudo-population}
\label{sec:holmberg}

Following  the popular Horvitz-Thompson (HT) approach to $\pi$ps sampling and estimation,   each unit $i \in \sample$, should be ``predicted'' in  ${\ \mathcal U^*_{N^*}}$ a number of times equal to its design weight   $\pi_i^{-1}$, assumed all integers. For the general non-integer case  the following strategy has been proposed since the 90s (\cite{holmberg:98}). Let $r_i = \pi_i^{-1} -  \lfloor \pi_i^{-1} \rfloor$, and consider independent Bernoulli r.v.s $\epsilon_i$s with $Pr ( \epsilon_i =1 \, \vert \DN , \, \YN , \, \XN ) = r_i$.  A HT pseudo-population  is constructed by replicating every sampled unit $i \in  \s$  $N^{*HT}_i = \lfloor \pi_i^{-1} \rfloor + \epsilon_i$ times, with corresponding values $y_i$, $x_i$. The size of a HT pseudo-population $N^{*HT} = \sum_{i=1}^{N} N^{*HT}_i D_i $ is in general  {\em not} equal to $N$. However, the ratio $N^{*HT} /N$ tends in probability to $1$ as  $N$, $n$ increase. Furthermore, and more strongly, it is easy to see that HT pseudo-population satisfies the regularity conditions $P1$-$P3$, and hence the resampling distribution of $\sqrt{n} ( \theta ( \widehat{F}^{*}_H ) - \theta ( F^{*}_{N^*}))$ tends to the same limit as the sampling distribution of $\sqrt{n} ( \theta ( \widehat{F}_H ) - \theta ( F_N  ))$.

%For each unit $i \in \sample$, let $r_i = \pi_i^{-1} -  \lfloor \pi_i^{-1} \rfloor$, and consider independentBernoulli r.v.s $\epsilon_i$s with $Pr ( \epsilon_i =1 \, \vert \DN , \, \YN , \, \XN ) = r_i$.Holmberg's proposal (cfr. \cite{holmberg:98}) consists in constructing a pseudo-population where every sampled unit $i \in  \s$is replicated $N^{*HOL}_i = \lfloor \pi_i^{-1} \rfloor + \epsilon_i$ times, with corresponding values $y_i$, $x_i$.The size of the pseudo-population$N^{*HOL} = \sum_{i=1}^{N} N^{*HOL}_i D_i $ is in general  {\em not} equal to $N$. However, the ratio $N^{*HOL} /N$ tends in probability to $1$ as  $N$, $n$ increase. Furthermore, and more strongly, it is easy to see that Holmberg pseudo-population satisfies the regularity conditions $P1$-$P3$, and hence the resampling distribution of $\sqrt{n} ( \theta ( \widehat{F}^{*}_H ) - \theta ( F^{*}_{N^*}))$ tends to the same limit as the sampling distribution of $\sqrt{n} ( \theta ( \widehat{F}_H ) - \theta ( F_N  ))$.  {\color{purple} {\bf si potrebbe far notare che: }  in the special case of  $r_i$ all null for every sample unit,  i.e. $N_i^{*HOL} = \pi_i^{-1}$, then the psudo-population results calibrated  to the  total (mean) of the auxiliary variable  in that  $ \sum_{i=1}^{N} N^{*HOL}_i  x_i D_i = \sum_{i=1}^{N}  x_i $}

%Notice that only in the special case of  $r_i$ all null for every sample unit,  i.e. $N_i^{*HT} = \pi_i^{-1}$, then the HT psudo-population results calibrated  to the mean of the auxiliary variable in that  $ \sum_{i=1}^{N} N^{*HT}_i  x_i D_i = \sum_{i=1}^{N}  x_i $

\subsection{Multinomial pseudo-population}
\label{sec:multinomial}

For $k=1, \, \dots , \, N$, perform  independent trials consisting in choosing a unit from the original sample, where each unit $i$ is selected
 with probability
\begin{eqnarray}
\pi_i^{-1} / \sum_{j \in \s} \pi_j^{-1} = x_i^{-1} / \sum_{j \in \s} x_j^{-1} . \nonumber
\end{eqnarray}
\noindent If at trial $k$ unit $i$ is selected, unit $k$ of the pseudo-population will take values $y^*_k = y_i$ and $x^*_k = x_i$.
If $N^{*MUL}_i$, $i \in \s$, is the number of replications of unit $i$ in the pseudo-population, then
(conditionally on $\DN$, $\YN$, $\XN$) the r.v. $( N^*_i ; \; i \in \s )$ possesses a multinomial distribution, with
\begin{eqnarray}
E [ N^{*MUL}_i \, \vert \DN , \, \YN, \, \XN ] = N D_i \pi_i^{-1} / \sum_{j=1}^{N} D_j \pi_j^{-1} \label{eq:mean_miltinom} \\
V ( N^{*MUL}_i \, \vert \DN , \, \YN, \, \XN ) = N \left ( D_i \pi_i^{-1} / \sum_{j=1}^{N} D_j \pi_j^{-1} \right )
\left ( 1- D_i \pi_i^{-1} / \sum_{j=1}^{N} D_j \pi_j^{-1} \right )   \label{eq:var_miltinom} \\
C ( N^{*MUL}_i , \, N^{*MUL}_h \,  \vert \DN , \, \YN, \, \XN ) = - N D_i D_h \pi_i^{-1} \pi_h^{-1} /  \left ( \sum_{j=1}^{N} D_j \pi_j^{-1} \right )^2 ,  \; h \neq i . \label{eq:cov_miltinom}
\end{eqnarray}
This approach goes essentially back to \cite{pfeffermann:04}  and  guarantees by construction  a pseudo-population calibrated w.r.t. the population size.
%The number of units of the pseudo-population is equal to $N$, the size of the original population. In this sense, {\em the pseudo-population is calibrated w.r.t. the population size}.
Again, the multinomial pseudo-population satisfies conditions $P1$-$P3$, so that the resampling distribution of
$\sqrt{n} ( \theta ( \widehat{F}^{*}_H ) - \theta ( F^{*}_{N^*}  ))$
tends to the same limit as the sampling distribution of $\sqrt{n} ( \theta ( \widehat{F}_H ) - \theta ( F_N  ))$.

\subsection{Conditional Poisson pseudo-population}
\label{sec:cps_pseudopop}

The HT scheme in Section \ref{sec:holmberg} is essentially based on drawing a Poisson sample from $\s$, where unit $i \in \s$ does have
inclusion probability $r_i$. A simple idea to calibrate such scheme in order to produce a pseudo-population of exactly $N$ units, consists in
defining the quantities
\begin{eqnarray}
\tau_i = N \frac{\pi_{i}^{-1}}{\sum_{k \in \s} \pi_k^{-1}} -
\left \lfloor N \frac{\pi_{i}^{-1}}{\sum_{k \in \s} \pi_k^{-1}} \right \rfloor
 , \;\; i \in \s
\nonumber
\end{eqnarray}
\noindent and in drawing from $\s$ a sample $\s_0$ of
\begin{eqnarray}
n_0 = \sum_{i \in \s} \tau_i = N - \sum_{i \in \s}
\left \lfloor N \frac{\pi_{i}^{-1}}{\sum_{k \in \s} \pi_k^{-1}} \right \rfloor
\nonumber
\end{eqnarray}
\noindent units, according to a conditional Poisson sampling design with first order inclusion probabilities $\tau_i$s.

For each unit $i \in \s$, let $\epsilon_i$ be equal to $1$ iff $i$ is in $\s_0$, and $\epsilon_i =0$ otherwise. Each unit $i$ of
the original sample is replicated in the pseudo-population exactly
\begin{eqnarray}
N^{*CPP}_i = \left \lfloor N \frac{\pi_{i}^{-1}}{\sum_{k \in \s} \pi_k^{-1}} \right \rfloor + \epsilon_i
\label{eq:N*_CPP}
\end{eqnarray}
\noindent times.

\begin{proposition}
\label{prop_NCPP}
The conditional Poisson pseudo-population satisfies conditions $P1$-$P3$.
\end{proposition}
%\vspace{-0.5cm}\noindent (Proof in Appendix)

As a consequence of Proposition \ref{prop_NCPP}, the resampling distribution of $\sqrt{n} ( \theta ( \widehat{F}^{*}_H ) - \theta ( F^{*}_{N^*}  ))$
tends to the same limit as the sampling distribution of $\sqrt{n} ( \theta ( \widehat{F}_H ) - \theta ( F_N  ))$.

\subsection{Double-Calibrated pseudo-population}
\label{sec:cal_pseudopop}

The conditional Poisson pseudo-population illustrated in Subsection \ref{sec:cps_pseudopop} is calibrated w.r.t. the population size $N$, but not w.r.t. the mean of
the auxiliary variable ${\mathcal X}$.
%since the mean of  ${\mathcal X}$ in the pseudo-population is not equal to the mean of  ${\mathcal X}$ in the original population.
A natural idea would consist in modifying $N^{*CPP}_i$ defined by equation ($\ref{eq:N*_CPP}$)  in order to satisfy a further constraint:
{\em the mean of ${\mathcal X}$ in the pseudo-population is equal to the mean of ${\mathcal X}$ in the actual population}.

%A fairly simple approach to reach the goal above consists in resorting to calibration.

Take $N^{*CPP}_i$, $i \in \s$ as an ``initial''
solution for replicates of sample units in the pseudo-population, and let further
\begin{eqnarray}
\overline{X}_N  = N^{-1} \sum_{i=1}^{N} x_i , \;\;
\overline{X}^*_{N^*} = N^{*-1} \sum_{i=1}^{N} N^*_i x_i D_i , \;\;
N^*  =  \sum_{i=1}^{N} N^*_i  D_i .
\end{eqnarray}

The basic idea is to choose pseudo-population replicates that satisfy both constraints on population size and mean of ${\mathcal X}$,
and that are as close as possible to the initial $N^{*CPP}_i$s. More formally, the pseudo-population replicates are
taken equal to $N^{*DCal}_i$s, the solution of
the following quadratic problem:
\begin{eqnarray}
\ipotc{{\mathrm{min}}  \;\; \sum_{i \in \s} (N^*_i -N^{*CPP}_i)^2}{N^* =N \;\;\;\; \;\;\;\; \;\;\;\;
\;\;\;\; \;\;\;\; \;\;\;\; \;\;\;\;}{\overline{X}^*_{N^*} = \overline{X}_N \;\;\;\; \;\;\;\; \;\;\;\; \;\;\;\; \;\;\;\; \;\;\;\;}{N^*_i
\ge 1 \;\;\;\; \;\;\;\; \;\;\;\; \;\;\;\; \;\;\;\; \;\;\;\; \;\;\;\; \;}
\label{eq:probl_min}
\end{eqnarray}

\begin{proposition}
\label{prop_calib}
The calibrated pseudo-population with replicates $N^{*DCal}_i$ that solves the optimization problem $( \ref{eq:probl_min} )$ possesses the following
property:
\begin{eqnarray}
\frac{N^{*DCal}_i}{N^{*CPP}_i} \stackrel{p}{\rightarrow} 1  \;\; {\mathrm{as}} \; n, \; N \, \rightarrow \infty .
\label{eq:conv_nstar}
\end{eqnarray}
\end{proposition}
%\vspace{-0.5cm}\noindent (Proof in Appendix)

Intuitively speaking, Proposition  \ref{prop_calib} tells us that as $N$, $n$ increase, the solution of the
optimization problem ($\ref{eq:probl_min} )$ tends to coincide with $N^{*CPP}_i$. Hence, for
``very large'' population and sample size, $ N^{*CPP}_i$s can be taken as a good approximation
of the actual solution of the optimization problem $( \ref{eq:probl_min} )$. Of course, this is only an asymptotic
result, and for the use of ``not too large'' $n$, $N$, the use of $N^{*DCal}_i$ instead of
$N^{*CPP}_i$ could produce considerably different results in the resampling procedure.

The values $N^{*DCal}_i$s obtained by solving $(\ref{eq:probl_min} )$ are not necessarily integer-valued.
In order to obtain integer values, it is enough to apply to $N^{*DCal}_i$s a randomization device similar to that of
CPP pseudo-population described in Section \ref{sec:cps_pseudopop}.

\subsection{Hot-deck pseudo-population}
\label{sec:htd_pseudopop}

The basic idea of the calibrated pseudo-population introduced in Subsection \ref{sec:cal_pseudopop} consists in constructing a
pseudo-population that is ``similar'' for some characteristics of the auxiliary variable ${\mathcal X}$ w.r.t. the original population.
This idea is pursued by taking only the sample $x_i$s values. Although in many practical cases this is true (for instance, when
the data user is different from the sample design planner, and only sample weights $\pi_i$s are available), in some
cases $x_i$s are available for all population units.
When all $x_i$s are available, the notion of (finite) population predictor can be extended by considering predictors of the form
$\{ ( x^{*}_i , \, y^{*}_i ), \; k=1, \, \dots , \, N  \}$, where $x^{*}_i =x_i$ for every unit $i=1, \, \dots , \, N$ and
$y^{*}_i = \widehat{y}_i =$ imputed value for $y_i$, according to hot-deck imputation.
%The imputation task for $y$ values can be accomplished by resorting to hot-deck imputation.
In detail, the hot-deck pseudo-population is composed by $N$ units, i.e. $\mathcal{U}^*_{N^*}={\mathcal U}^*_N$.
A pair of values $(x^*_i , \, y^*_i )$ corresponds to each unit $i \in {\mathcal U}^*_N$, with
\begin{eqnarray}
x^*_i & = & x_i , \;\; i= 1, \, \dots , \, N \label{eq:hotdeckx} \\
y^*_i & = & \ipot{y_{i} \;\; {\mathrm{if}} \; i \in \s \;\;\;\; \;\;\;\; \;\;\;\; \;\;\;\; \;\;\;\; \;\;\;\;
\;\;\;\; \;\;\;\; \;\;\;\; \;\;\;\; \;\;\;\; \;\;\;\; \;\;\;\; \;\;}{y_j \; {\mathrm{with}} \; j = {\mathrm{argmin}}_{j \in \s} \left \vert
x_j - x_i  \right \vert \; {\mathrm{if}} \; i \in {\mathcal U}^*_N \setminus \s} . \label{eq:hotdecky}
\end{eqnarray}

In other terms, for each unit $i \in {\mathcal U}^*_N$ a {\em donor unit}  $j(i)$ is chosen, such that
\begin{eqnarray}
j(i) := \ipot{i \; {\mathrm{if}} \; i \in \s \;\;\;\; \;\;\;\; \;\;\;\; \;\;\;\; \;\;\;\; \;\;\;\;
\;\;\;\; \;\;\;\; \;\;\;\; \;\;\;\; \;\;\;\; \;\;\;\; \;\;\;\; \;\;\;\; \;\;\;\;}{\left \vert x_{j(i)} - x_i \right \vert
= \min_{j \in U_{N} \setminus \s}  \left \vert
x_j - x_i  \right \vert \; {\mathrm{if}} \; i \in {\mathcal U}^*_N \setminus \s .} \nonumber
\end{eqnarray}
\noindent The values $x^*_i$, $y^*_i$ for unit $i$ are then taken equal to those of its donor, leading to a pseudo-population which is calibrated by construction w.r.t.  both population size $N$ and the entire distribution of the auxiliary variable ${\mathcal X}$.

\begin{proposition}
\label{prop_hotdeck}
If the pseudo-population is constructed {\em via} hot-deck imputation of $y$s values, then, as $n$, $N$ increase,
the resampling distribution of $\sqrt{n} ( \theta ( \widehat{F}^{*}_H ) - \theta ( F^{*}_{N^*}  ))$
tends to the same limit as the sampling distribution of $\sqrt{n} ( \theta ( \widehat{F}_H ) - \theta ( F_N  ))$.
\end{proposition}
%\vspace{-0.5cm}\noindent (Proof in Appendix)

\section{Simulation Study}
\label{sec:simul}

Main goal of the simulation is to  empirically evaluate  the effects that  different choices for constructing the pseudo-population  ${\mathcal U}^*_{N^*}$ (where  resampling is actually performed) may have  upon the  accuracy of the resulting inference in practical applications. The  simulation has been designed by focusing  three key points:
\begin{itemize}
\item[$a)$] exploration of small to moderate $n$ and $N$ in order to highlight differences due to finite sizes as well as to evaluate approximations based on asymptotic arguments as given in the first part of the present paper;
\item[$b)$] analysis of specific features of the  pseudo-population ${\mathcal U}^*_{N^*}$ due to different construction choices;
\item[$c)$] investigation of the statistical properties of the final  estimates provided  by resampling  into such different pseudo-populations.
\end{itemize}
The simulated scenarios, parameters and estimators  are summarized in Table $\ref{tab:tab1}$.
For the sake of comparisons, beside the five strategies proposed in Section \ref{sec:calib_1},  the direct bootstrap (\cite{antal:11})  is also simulated, since  it is a recent competitor based on a non-predictive resampling approach.  The  variates ${\mathcal Y}$, ${\mathcal X}$ have  been simulated under the same model as in \cite{antal:11}.

\bigskip
{\small
\begin{table}[htbp]
\caption{\small Simulated scenarios, population parameters and estimators}
\label{tab:tab1}
{\small
\begin{tabular}{lll}
\hline \hline
\textit{Scenarios} & & \\
$N=200, \ {\bf 400}$ & &  $n=(0.2 N)= 40, \ {\bf 80}$ \\ \smallskip
correlation between  ${\mathcal Y}$ and ${\mathcal X}$ & $\simeq 0.8$ & \\
\hline\hline
\textit{Parameters}  & & \textit{H\'{a}jek Estimators}   \\ \smallskip
 $\bar{Y}_N=\sum_{i=1}^N y_i/N$  & & $\hat{\bar Y}_H=\sum_{i=1}^N D_i \pi^{-1}_i y_i/\sum_{i=1}^N D_i \pi^{-1}_i$  \\
$Q_N(p)=\inf \{y:F_N(y) \ge p\}$   &&   $\hat Q_H(p)=\inf\{y:\hat F_H(y) \ge p\}  \:\: y \in \mathbb{R}$\\
with $p=0.5, 0.75$     & &  $ \, $ \\
\hline \hline
\end{tabular}
}
\end{table}
}

Samples have been simulated under two different fixed size  $\pi$ps designs of increasing entropy:   Pareto sampling and   (normalized) conditional Poisson sampling (CPS for short), this latter  already mentioned in Section \ref{sec:assumptions_prel} as a maximum entropy design. Notice that Pareto design is high entropy, although not yet proved  asymptotically maximum entropy; however it is heuristically recognized to be  very close to the asymptotically maximum entropy Rao-Sampford design  (\cite{bondesson06}). Moreover, unlike the CPS design, the Pareto sampling is very simple to implement, and can be used in simple  acceptance-rejection rules to produce CPS samples with a significant reduction  of computational burden. Simulation has been implemented partly in {\em {Mathematica}} code  and partly in the R environment.  $1000$ Monte Carlo ($MC$) runs,   simulating the sample space, have been combined with $M=1000$ resampling runs from each generated sample. The $MC$ error deriving from these choices has been controlled via the empirical bias  of the (unbiased) Horvitz-Thompson estimator $\hat{\bar Y}_{HT}$,  and it has been kept under 1\% (relative to the true population mean  $\bar Y$).

Simulation results are gathered in Tables 2-5 where the simulated methods to construct the pseudo-population are indicated by the following  acronyms:  $HT$ illustrated in subsection \ref{sec:holmberg}; $MUL$  for the Multinomial pseudo-population in \ref{sec:multinomial}; $CPP$ for the conditional Poisson pseudo-population in \ref{sec:cps_pseudopop}; $DCal$ for the double-calibrated pseudo-population in \ref{sec:cal_pseudopop};  $HD$ for the hot-deck pseudo-population  in \ref{sec:htd_pseudopop}; and   $Dir$  for the (non-predictive, non pseudo-population based) direct bootstrap  (\cite{antal:11}).

\bigskip
\noindent *****************TABLE 2  ABOUT HERE ************************
\bigskip

Results in Table $\ref{tab:tab2}$  offer indications about the ability of the pseudo-population ${\mathcal U}^*_{N^*}$  as a {\em{predictor} }of the actual population ${\mathcal U}_N$, according to key point $b)$ above.  Except for the direct bootstrap involving no pseudo-population, it has been checked in two respects:  $i)$  the pseudo-population size $N^*$ and mean of the auxiliary variable  $\bar X^*$ as {\em predictors} of (known) population $N$ and $\bar X_{N}$ respectively, as measured  via empirical (relative) bias  $RB\left[N^*; N\right] =  100 \times \left[E_{MC}(N^*) - N\right]/N$ (where $E_{MC}$ indicates the average over all the Monte Carlo runs and $RB\left[\bar X^*; \bar X_{N}\right]$ follows accordingly); and  $ii)$ how able the pseudo-population is to reproduce the actual  p.d.f.  as measured by  the maximal MC value of the  Kolmogorov statistic $\max_{MC} \sup_{y} \left| F^*_{N^*}(y) - F_N(y)\right|,  \; y \in \mathbb{R}. $

A clear connection appears between the conservation of both  $N$ and $\bar X$ and the ability of reproducing the entire population d.f.: indeed $HD$ and $DCal$ pseudo-populations, which account for such a conservation to the largest extent, emerge as the best performers, uniformly in all the simulated scenarios.
Also, this reflects on the ability of the resampling algorithm based on such pseudo-populations,  to reproduce the estimator distribution.
 %and ultimately to provide ``good'' $p$-values.

\bigskip
\noindent *****************TABLEs 3 \& 4  ABOUT HERE ************************
\bigskip

According to key point $c)$ above,  both kinds of confidence intervals (CI)  ilustrated in section \ref{sec:resampling_1} have been simulated.  Table \ref{tab:tab3} concerns  CI (\ref{conf_int_percentile})  which basically correspond to  {\em  bootstrap percentile} method, and  Table \ref{tab:tab4} refers to CI  (\ref{eq:conf_int_norm}).
Performances at (nominal) confidence level $95$\% has been  investigated  {\em via}  empirical coverage (Cov), with respect to the true population parameter,  and average length (AL).
Notice  that although the {\em percentile} method is the crudest available for producing CI via resampling, we rate it  appropriate for the goals of the present simulation  because it  allows the evaluation of  the ability of the resampling algorithm to produce $p$-values, and ultimately  to reproduce the estimator sampling distribution particularly in its tails.
In Table \ref{tab:tab3} all the methods investigated for constructing ${\mathcal U}^*_{N^*}$ provide acceptable levels of empirical coverage based on the $0.025$ and $0.975$ percentiles of the resampling distribution. Moreover they  all  tend to improve for increasing sizes $N$ and $n$, as expected according  to asymptotic results in Section \ref{sec:resampling_1}. However $HD$ and $DCal$, which provide the best predictor of ${\mathcal U}_N$, also give the best  coverage probabilities, uniformly in all scenarios simulated  for both linear and non linear estimators. Notably, $HD$ shows the largest average lengths  in addition to the largest empirical coverages, which suggests a tendency  to  supply  conservative CI.

A similar behaviour can be observed  in Table $\ref{tab:tab4}$,  although the resampling plays here a minor role, limited to the (point) bootstrap estimate (\ref{eq:resamp_var}) for the estimator variance then coupled with  standard normal distribution percentiles.  Notice that this is also  the method for interval estimation suggested for the {\em{non-predictive}} direct bootstrap.  However,  $Dir$ exhibits lower empirical coverage probabilities than the  {\em{predictive}} pseudo-population based  methods, seemingly  due to systematic smaller lengths. The notable exception of  $DCal$  may be explained by its weaker ability to produce accurate point bootstrap estimates than the other {\em{predictive}} methods simulated. Still $HD$ emerges as the best performer for uniformly giving the larger empirical coverages in all scenarios simulated and for maintaining its conservative peculiarity.

Finally  and as a desirable feature of a resampling  algorithm applying to  complex sampling  from finite populations, it has been investigated a popular property of the  classic {\em i.i.d.} Efron's bootstrap:  the ability of the resampled distribution of an estimator of the population mean to match the (original) sample mean as its empirical first moment. Such property, dubbed  {\em bootstrap unbiasedness}, has been  measured   by  the  (percentage) relative bias
$RB\left[\hat\theta^*_m; \hat\theta \right] =100 \times E_{MC} \left\{ \left[E^*( \hat\theta^*_m) -  \hat\theta\right]/\hat\theta\right\}$
where $E^*$ indicates the empirical average over the $M$ resampling runs and by taking $\hat \theta=\bar Y$ and   $\hat\theta^*_m, m=1 \cdots M$ as its resampled  distribution.    Table $\ref{tab:tab5}$  reports simulation results with respect to both Horvitz-Thompson and H\'ajek estimation of population mean.  Empirical evidence confirms $HT$ and $Dir$ as  algorithms purposively constructed under the conventional Horvitz-Thompson paradigm for linear parameters. All the other proposed strategies for producing  the pseudo-population  perform well under the H\'ajek approach to estimation. Again $HD$ appears as ensuring bootstrap-unbiasedness to the largest extent.

\bigskip
\noindent *****************TABLE 5   ABOUT HERE ************************
\bigskip

As a  final remark concerning the actual implementation of specific algorithms, note that
 all the simulated populations have been checked to ensure $\pi_i < 1, \quad i=1 , \, \dots , \, N$. However, for $MUL$   it may still  occur  $\pi^*_k \ge 1$ for one or more (sampled) unit $k$ included in the pseudo-population. This empirically appears to be often the case as the number of MC runs increases.  As a consequence, an {\em ad hoc} routine has to be implemented on top of the resampling algorithm, aiming at including  such units in each bootstrap sample and sequentially recomputing the resampling inclusion probability until they are all strictly smaller than 1,  and by simultaneously reducing the (re)sample size accordingly (see, for instance, \cite{tille06} for details).

  %$ii)$ The ({\em{non-predictive}} ) $Dir$  algorithm involves a rather complex resampling scheme that does not mimick the original sampling design by combining instead three different designs. This would affect to an uncontrolled extent the resampling inclusion probability $\pi^*_k$s accordingly. However, according to \cite{tille06}, the original inclusion probabilities $\pi_i$s for all re-sampled units are in fact fed to the algorithm  to compute estimate replications and ultimately to produce the bootstrap distribution. And $iii)$  all the resampling algorithms simulated are significant resource consumers; however, the five {\em predictive} strategies are essentially equivalent w.r.t the computational burden, while the {\em non-predictive, non pesudo-population based } competitor $Direct$ bootstrap  has comparatively required  up to five times computational time.   \\

\section {Conclusions}
\label{sec:conclus}

In this paper a new class of resampling methods applying to non-{\em i.i.d.} finite population sampling is proposed under a principled  {\em predictive} approach.  The proposed resampling unifies any  method based on pseudo-populations, i.e. according to the {\em plug-in} principle upon which the original Efron's bootstrap is based.   A large sample theory is derived for the predictive resampling, in the  H\'ajek finite population asymptotic setup,  and in the same spirit of the  classical asymptotics  for {\em i.i.d.} bootstrap by  \cite{bickelfried81}.  It is also  proved that all techniques to produce  the pseudo-population  are asymptotically equivalent, under mild regularity conditions.

In addition, five strategies  have been illustrated  for constructing the pseudo-population in practice.  Two of them go back to results already appeared in the literature and the remaining three are new proposals with improved performance, as shown in a simulation study. Empirical evidence  confirms  that how to construct the pseudo-population is a crucial choice for small to moderate population and sample sizes, under general  sampling designs such as  $\pi$ps designs.  As a general recommendation such choice should be guided by enforcing the ability of the pseudo-population  to be a {\em good predictor} of the actual population.  The simulation study indicates  the pseudo-population based on hot-deck imputation ($HD$) as the soundest method,  provided that auxiliary $x_i$s values are available for all population units. When $x_i$s are known only for sample units, as it might be the case in applications,  good results are offered  by a pseudo-population calibrated w.r.t. both the population size and the mean (total) of the auxiliary variable ($DCal$), when combined with percentile confidence intervals.

\newpage
%***********************TABLES********************

{\small
\begin{table}[htbp]
\caption{{\small ${\mathcal U}^*_{N^*}$ as a predictor of ${\mathcal U}_N \ \ (N=200 \quad {\bf{400}})$}}
\label{tab:tab2}
{\small
\begin{tabular}{cccc}
\hline \hline
 & $RB\left[\bar X^*; \bar X_{N}\right]$ & $RB\left[N^*; N\right]$  & $Sup_{MC} \left| F^*_{N^*}(y) - F_N(y)\right|$ \\
 {\small  PARETO sampling design} & & & \\
 $HT$ &   0,03  \quad  {\bf{0,04}} &  -0,44  \quad {\bf{0,38}} & 0,87	  \quad {\bf {0,51}} \\
 $MUL$ &   5,46  \quad  {\bf{3,39}} &  0  \quad {\bf{0}} & 0,93  \quad {\bf {0,54}} \\
 $CPP$ &   5,46 \quad  {\bf{3,38}} &  0  \quad {\bf{0}} & 0,88	  \quad {\bf {0,52}} \\
 $DCal$ &  0,02 \quad {\bf{-0,02}} &  0,003  \quad {\bf{-0,01}} & 0,55	  \quad {\bf {0,46}} \\
 $HD$   &    0     \quad  {\bf{0}} &  0  \quad {\bf{0}} & 0,47	  \quad {\bf {0,37}} \\
  \hline
{\small   CPS sampling design} & & &\\
 $HT$ &   -0,02  \quad  {\bf{0,02}} &  -1,05  \quad {\bf{-1,40}} & 0,50	  \quad {\bf {0,52}} \\
 $MUL$ &   5,06  \quad  {\bf{3,39}} &  0  \quad {\bf{0}} & 0,53  \quad {\bf {0,55}} \\
 $CPP$ &   5,04 \quad  {\bf{3,88}} &  0  \quad {\bf{0}} & 0,51	  \quad {\bf {0,52}} \\
 $DCal$ &  0,06 \quad {\bf{-0,04}} &  0,04  \quad {\bf{-0,04}} & 0,46	  \quad {\bf {0,47}} \\
 $HD$   &    0     \quad  {\bf{0}} &  0  \quad {\bf{0}} & 0,48	  \quad {\bf {0,33}} \\
  \hline \hline
 \end{tabular}
}
\end{table}
}

{\small
\begin{table}[htbp]
\caption{{\small $95\%$ Resampling CI -   $percentile$ method \ \ $(N=200 \quad {\bf{400}})$}}\label{tab:tab3}
{\small
\begin{tabular}{ccccccc}
\hline \hline
{ \small PARETO}  &\multicolumn{2}{c}{$\hat{\bar{Y}}_H$} & \multicolumn{2}{c}{$\hat{Q}_N{0.5}$} & \multicolumn{2}{c}{$\hat{Q}_H(0.75)$ } \\
   & $Cov$  & $AL$ & $Cov$ & $AL$ &  $Cov$ & $AL$ \\
$HT$ &	0,89	\ {\bf{	0,90	}} &	0,23	\ {\bf{	0,17	}} &	0,88	\ {\bf{	0,91	}} &	0,33	\ {\bf{	0,22	}} &	0,91	\ {\bf{	0,93	}} &	0,37	 \ {\bf{	0,28	 }}\\
$MUL$ &	0,87	\ {\bf{	0,89	}} &	0,23	\ {\bf{	0,02	}} &	0,73	\ {\bf{	0,79	}} &	0,33	\ {\bf{	0,03	}} &	0,82	\ {\bf{	0,79	}} &	0,38	 \ {\bf{	0,03	 }}\\
$CPP$ &	0,89	\ {\bf{	0,90	}} &	0,23	\ {\bf{	0,17	}} &	0,89	\ {\bf{	0,92	}} &	0,33	\ {\bf{	0,22	}} &	0,92	\ {\bf{	0,94	}} &	0,38	 \ {\bf{	0,29	 }}\\
$DCal$ &	0,95	\ {\bf{	0,95	}} &	0,24	\ {\bf{	0,18	}} &	0,95	\ {\bf{	0,97	}} &	0,33	\ {\bf{	0,23	}} &	0,95	\ {\bf{	0,95	 }} &	 0,39	\ {\bf{	0,30	 }}\\
$HD$ &	0,97	\ {\bf{	0,98	}} &	0,27	\ {\bf{	0,20	}} &	0,95	\ {\bf{	0,95	}} &	0,36	\ {\bf{	0,26	}} &	0,99	\ {\bf{	0,99	}} &	0,43	 \ {\bf{	0,33	 }}\\
\hline
 { \small   CPS } & & &\\ \\
$HT$ &	0,90	\ {\bf{	0,91	}} &	0,24	\ {\bf{	0,17	}} &	0,91	\ {\bf{	0,92	}} &	0,33	\ {\bf{	0,22	}} &	0,90	\ {\bf{	0,93	}} &	0,38	 \ {\bf{	0,29	 }}\\
$MUL$ &	0,89	\ {\bf{	0,92	}} &	0,24	\ {\bf{	0,17	}} &	0,73	\ {\bf{	0,81	}} &	0,34	\ {\bf{	0,22	}} &	0,82	\ {\bf{	0,79	}} &	0,39	 \ {\bf{	0,29	 }}\\
$CPP$ &	0,90	\ {\bf{	0,90	}} &	0,24	\ {\bf{	0,17	}} &	0,91	\ {\bf{	0,92	}} &	0,34	\ {\bf{	0,22	}} &	0,91	\ {\bf{	0,94	}} &	0,38	 \ {\bf{	0,29	 }}\\
$DCal$ &	0,96	\ {\bf{	0,96	}} &	0,25	\ {\bf{	0,18	}} &	0,98	\ {\bf{	0,97	}} &	0,34	\ {\bf{	0,23	}} &	0,95	\ {\bf{	0,94	 }} &	 0,40	\ {\bf{	0,30	 }}\\
$HD$ &	0,98	\ {\bf{	0,98	}} &	0,27	\ {\bf{	0,20	}} &	0,96	\ {\bf{	0,95	}} &	0,37	\ {\bf{	0,26	}} &	0,99	\ {\bf{	0,99	}} &	0,44	 \ {\bf{	0,33	 }}\\
\hline\hline
 \end{tabular}
}
\end{table}
}

{\small
\begin{table}[htbp]
\caption{{\small $95\%$ Standard Normal CI  with resampling variance estimate  $\ \ (N=200 \quad {\bf{400}})$}}
\label{tab:tab4}
{\small
\begin{tabular}{ccccccc}
\hline \hline
{ \small PARETO}  &\multicolumn{2}{c}{$\hat{\bar{Y}}_H$} & \multicolumn{2}{c}{$\hat{Q}_N{0.5}$} & \multicolumn{2}{c}{$\hat{Q}_H(0.75)$ } \\
   & $Cov$  & $AL$ & $Cov$ & $AL$ &  $Cov$ & $AL$ \\
$HT$ &	0,90	\ {\bf {	0,91	}} &	0,24	\ {\bf {	0,17	}} &	0,90	\ {\bf {	0,91	}} &	0,36	\ {\bf {	0,24	}} &	0,93	\ {\bf {	 0,91	}} &	 0,40	\ {\bf {	0,30	}}\\
$MUL$ &	0,90	\ {\bf {	0,91	}} &	0,24	\ {\bf {	0,17	}} &	0,89	\ {\bf {	0,92	}} &	0,36	\ {\bf {	0,24	}} &	0,92	\ {\bf {	 0,91	}} &	 0,41	\ {\bf {	0,30	}}\\
$CPP$ &	0,91	\ {\bf {	0,92	}} &	0,24	\ {\bf {	0,17	}} &	0,89	\ {\bf {	0,92	}} &	0,36	\ {\bf {	0,24	}} &	0,93	\ {\bf {	 0,92	}} &	 0,40	\ {\bf {	0,30	}}\\
$DCal$ &	0,84	\ {\bf {	0,86	}} &	0,25	\ {\bf {	0,18	}} &	0,85	\ {\bf {	0,89	}} &	0,38	\ {\bf {	0,25	}} &	0,88	 \ {\bf {	 0,90	}} &	 0,43	\ {\bf {	0,33	}}\\
$HD$ &	0,91	\ {\bf {	0,93	}} &	0,27	\ {\bf {	0,20	}} &	0,92	\ {\bf {	0,94	}} &	0,40	\ {\bf {	0,27	}} &	0,95	\ {\bf {	 0,96	}} &	 0,44	\ {\bf {	0,34	}}\\
$Dir$ &	0,89	\ {\bf {	0,90	}} &	0,22	\ {\bf {	0,16	}} &	0,86	\ {\bf {	0,87	}} &	0,32	\ {\bf {	0,21	}} &	0,92	\ {\bf {	 0,90	}} &	 0,38	\ {\bf {	0,28	}}\\ \\
\hline
 { \small   CPS } & & &\\ \\
$HT$ &	0,91	\ {\bf {	0,91	}} &	0,24	\ {\bf {	0,17	}} &	0,89	\ {\bf {	0,90	}} &	0,37	\ {\bf {	0,24	}} &	0,92	\ {\bf {	 0,90	}} &	 0,40	\ {\bf {	0,30	}}\\
$MUL$ &	0,90	\ {\bf {	0,92	}} &	0,24	\ {\bf {	0,17	}} &	0,89	\ {\bf {	0,92	}} &	0,38	\ {\bf {	0,24	}} &	0,90	 \ {\bf {	 0,90	}} &	 0,41	\ {\bf {	0,31	}}\\
$CPP$ &	0,91	\ {\bf {	0,92	}} &	0,24	\ {\bf {	0,17	}} &	0,90	\ {\bf {	0,91	}} &	0,37	\ {\bf {	0,24	}} &	0,92	\ {\bf {	 0,91	}} &	 0,40	\ {\bf {	0,31	}}\\
$DCal$ &	0,85	\ {\bf {	0,87	}} &	0,25	\ {\bf {	0,19	}} &	0,87	\ {\bf {	0,87	}} &	0,39	\ {\bf {	0,25	}} &	0,89	 \ {\bf {	 0,89	}} &	 0,44	\ {\bf {	0,33	}}\\
$HD$ &	0,94	\ {\bf {	0,95	}} &	0,27	\ {\bf {	0,20	}} &	0,90	\ {\bf {	0,93	}} &	0,40	\ {\bf {	0,27	}} &	0,97	\ {\bf {	 0,95	}} &	 0,45	\ {\bf {	0,34	}}\\
$Dir$ &	0,90	\ {\bf {	0,90	}} &	0,23	\ {\bf {	0,16	}} &	0,85	\ {\bf {	0,88	}} &	0,33	\ {\bf {	0,21	}} &	0,92	\ {\bf {	 0,88	}} &	 0,38	\ {\bf {	0,28	}}\\
\hline\hline
 \end{tabular}
}
\end{table}
}

{\small
\begin{table}[htbp]
\caption{{\small Bootstrap-unbiasedness $\ \ (N=200 \quad {\bf{400}})$}}
\label{tab:tab5}
{\small
\begin{tabular}{cccccc}
\hline\hline
 & { \small PARETO} &&  & { \small   CPS } & \\
 & $RB\left[{\hat{\bar Y}}^*_{HT} -  {\hat{\bar Y}}_{HT} \right]$ &  $RB\left[{\hat{\bar Y}}^*_{H} -  {\hat{\bar Y}}_{H} \right]$  &   & $RB\left[{\hat{\bar Y}}^*_{HT} -  {\hat{\bar Y}}_{HT} \right]$ &  $RB\left[{\hat{\bar Y}}^*_{H} -  {\hat{\bar Y}}_{H} \right]$ \\
$HT$ &	0,06	\ {\bf{	-0,16	}} &	0,87	\ {\bf{	0,72	}} & & -0,16	\ {\bf{	-0,13	}} &	0,97	\ {\bf{	0,63	}} \\
$MUL$ &	5,57	\ {\bf{	3,11	}} &	0,92	\ {\bf{	0,65	}}  & & 4,96	\ {\bf{	3,74	}} &	1,07	\ {\bf{	0,65	}} \\
$CPP$ &	5,46	\ {\bf{	3,15	}} &	0,84	\ {\bf{	0,70	}} & & 4,84	\ {\bf{	3,73	}} &	1,01	\ {\bf{	0,64	}} \\
$DCal$ &	1,66	\ {\bf{	1,12	}} &	-0,33	\ {\bf{	0,20	}} &  & 1,38	\ {\bf{	1,36	}} &	-0,34	\ {\bf{	-0,11	}} \\
$HD$ &	3,17	\ {\bf{	2,07	}} &	0,35	\ {\bf{	0,59	}} & & 2,55	\ {\bf{	2,27	}} &	0,34	\ {\bf{	0,28	}} \\
$Dir$ &	0,01	\ {\bf{	-0,01	}} &	0,70	\ {\bf{	0,41	}} & & -0,02	\ {\bf{	0,01	}} &	0,68	\ {\bf{	0,41	}} \\
\hline\hline
\end{tabular}
}
\end{table}
}

\newpage

\centerline{{\Large \noindent {\bf Appendix}}}

\begin{lemma}
\label{lemma_1}
Let $d_N = \sum_{i=1}^{N} \pi_i (1- \pi_i )$. Then, as $N$ increases,
\begin{eqnarray}
\frac{d_N }{N} \rightarrow d = f \left ( 1- \frac{\mathbb{E} [ X_1^2 ]}{\mathbb{E} [ X_1 ]^2} \right ) +
f (1-f) \frac{\mathbb{E} [ X_1^2 ]}{\mathbb{E} [ X_1 ]^2} \;\; a.s.- \mathbb{P} .
\label{eq:val_d}
\end{eqnarray}
\end{lemma}
\noindent
\begin{proof}[{\bf Proof of Lemma \ref{lemma_1}}]
Taking into account that $\pi_i= f_N x_i / \overline{X}_N$, with $f_N = n/N$ and $\overline{X}_{N} = \sum_{i=1}^{N} x_i /N$,
it is enough to observe that
\begin{eqnarray}
\frac{d_{N}}{N} & = &  \frac{1}{N} \sum_{i=1}^{N} \frac{f_{N}}{\overline{X}_{N}} x_i \left ( 1- \frac{f_{N}}{\overline{X}_{N}} x_i  \right ) \nonumber \\
\, & = & f_N - \left (  \frac{f_{N}}{\overline{X}_{N}}   \right )^2 \frac{1}{N} \sum_{i=1}^{N} x_i^2 \nonumber
\end{eqnarray}
\noindent and to apply the strong law of large numbers.
\end{proof}

\begin{lemma}
\label{lemma_2}
Consider the quantity $K_{\alpha} (y)$ in $( \ref{eq:def_K} )$.
The following results hold:
\begin{eqnarray}
\frac{1}{N} \sum_{i=1}^{N} \frac{1}{\pi_{i}} \left (  I_{( y_{i} \leq y )} - F_N (y) \right )
\rightarrow \frac{\mathbb{E} [ X_{1} ]}{f} \left ( K_{-1} (y) - \mathbb{E} \left [ X_1^{-1} \right ] \right ) F(y) \;\;
{\mathrm{as}} \; N \rightarrow \infty, \;\; a.s.- \mathbb{P} ;
\label{eq:limit_a} \\
\frac{1}{N} \sum_{i=1}^{N} ( 1- \pi_{i}) \left (  I_{( y_{i} \leq y )} - F_N (y) \right )
\rightarrow f F(y) \left ( 1- \frac{K_{1} (y)}{\mathbb{E} [ X_{1} ]}  \right )
 \;\;
{\mathrm{as}} \; N \rightarrow \infty, \;\; a.s.- \mathbb{P} .
\label{eq:limit_b}
\end{eqnarray}
\end{lemma}
\noindent
\begin{proof}[{\bf Proof of Lemma \ref{lemma_2}}]
Using the same notation as in Lemma $\ref{lemma_1}$, from $\pi_i= f_N x_i / \overline{X}_N$ it follows that
\begin{eqnarray}
\, & \, & \frac{1}{N} \sum_{i=1}^{N} \frac{1}{\pi_{i}} \left (  I_{( y_{i} \leq y )} - F_N (y) \right ) =
\frac{\overline{x}_{N}}{f_{N}} \left ( \frac{1}{N} \sum_{i=1}^{N} \frac{1}{x_{i}} I_{( y_{i} \leq y )} -
F_N (y) \frac{1}{N} \sum_{i=1}^{N} \frac{1}{x_{i}} \right ) \nonumber \\
\, & \, & \rightarrow  \frac{\mathbb{E} [ X_{1} ]}{f} \left ( \mathbb{E} \left [ X_1^{-1} I_{( Y_{1} \leq y )} \right ]
- \mathbb{E} \left [ X_1^{-1}  \right ] F(y)
\right ) \;\;
{\mathrm{as}} \; N \rightarrow \infty, \;\; a.s.- \mathbb{P} \nonumber
\end{eqnarray}
\noindent by the strong law of large numbers. Proof of $( \ref{eq:limit_a} )$ is completed by observing that
\begin{eqnarray}
\mathbb{E} \left [ X_1^{-1} I_{( Y_{1} \leq y )} \right ]  = F(y)  \mathbb{E} \left [ \left . X_1^{-1} \, \right \vert Y_{1} \leq y \right ]
= F(y) K_{-1} (y) .
\nonumber
\end{eqnarray}
Proof of $( \ref{eq:limit_b} )$ is similar.
\end{proof}

\begin{lemma}
\label{lemma_3}
Define the quantities
\begin{eqnarray}
Z_{i,N} (y) & = &  \left (  I_{( y_{i} \leq y )} - F_N (y) \right ) - \pi_i \frac{\sum_{i=1}^{N} (1- \pi_{i})
\left (  I_{( y_{i} \leq y )} - F_N (y) \right )}{\sum_{i=1}^{N} \pi_{i} (1- \pi_{i} )} , \;\; i=1, \, \dots , \, N ;
\label{eq:def_zin} \\
S^2_N (y) & = & \sum_{i=1}^{N} \left ( \frac{1}{\pi_{i}} -1 \right ) Z_{i,N} (y)^2 .
\label{eq:def_s2n}
\end{eqnarray}
Then, as $N$ goes to infinity, a.s.-$\mathbb{P}$, the following results hold
\begin{eqnarray}
& \, & Z_{i,N} (y) - \left (  I_{( y_{i} \leq y )} - F_N (y) \right ) \rightarrow
- \frac{f^{2}}{\mathbb{E} [ X_{1} ]} X_i \frac{\left ( 1- K_{1} (y) / \mathbb{E} [ X_1 ] \right)}{d} F(y)
 ;
\label{eq:limite_zin} \\
\frac{1}{N} S^2_N (y) & \rightarrow &
\left ( \frac{\mathbb{E} [ X_{1} ]}{f} K_{-1} (y) -1 \right ) F(y) (1- F(y) ) -
\frac{\mathbb{E} [ X_{1} ]}{f} \left ( K_{-1} (y) - \mathbb{E} [ X_{1}^{-1} ] \right ) F(y)^2 \nonumber \\
& - &
\frac{f^{2}}{d} \left ( 1- \frac{K_{1} (y)}{\mathbb{E} [ X_{1}]}  \right )^2 F(y)^2
.
\label{eq:limite_s2n}
\end{eqnarray}
\end{lemma}
\noindent
\begin{proof}[{\bf Proof of Lemma \ref{lemma_3}}]
Relationship $( \ref{eq:limite_zin}  )$ is an immediate consequence of Lemmas $\ref{lemma_1}$, $\ref{lemma_2}$. As far as $( \ref{eq:limite_s2n} )$
is concerned, observe first that
\begin{eqnarray}
\frac{S^{2}_{N}(y)}{N} = B_{1,N}(y) + B_{2,N}(y) + B_{3,N}(y) \label{eq:eq:scomponi_s2n}
\end{eqnarray}
\noindent where
\begin{eqnarray}
B_{1,N}(y) & = & \frac{1}{N} \sum_{i=1}^{N}  \left ( \frac{1}{\pi_{i}} -1 \right ) \left (  I_{( y_{i} \leq y )} - F_N (y) \right )^2 \nonumber \\
B_{2,N}(y) & = & \frac{1}{N} \sum_{i=1}^{N}  \pi_i  (  1 - \pi_{i} )  \left (
\frac{\frac{1}{N} \sum_{i=1}^{N} (1- \pi_i ) (  I_{( y_{i} \leq y )} - F_N (y) )}{\frac{1}{N} \sum_{i=1}^{N} \pi_i (1- \pi_i )}
\right )^2 \nonumber \\
B_{3,N}(y) & = & - \frac{2}{N} \sum_{i=1}^{N}   ( 1- \pi_{i} ) \left (  I_{( y_{i} \leq y )} - F_N (y) \right )
\frac{\frac{1}{N} \sum_{i=1}^{N} (1- \pi_i ) (  I_{( y_{i} \leq y )} - F_N (y) )}{\frac{1}{N} \sum_{i=1}^{N} \pi_i (1- \pi_i )} .
\nonumber
\end{eqnarray}
Next, it is not difficult to see that
\begin{eqnarray}
B_{1,N}(y) = \frac{1}{N} \sum_{i=1}^{N}  \left ( \frac{1}{\pi_{i}} -1 \right ) I_{( y_{i} \leq y)}
+  \frac{F_N (y)^2}{N} \sum_{i=1}^{N}  \left ( \frac{1}{\pi_{i}} -1 \right )
-  \frac{2 F_N (y)}{N} \sum_{i=1}^{N}  \left ( \frac{1}{\pi_{i}} -1 \right ) I_{( y_{i} \leq y)}
\label{eq:scomponi_b1n}
\end{eqnarray}
\noindent with
\begin{eqnarray}
\frac{1}{N} \sum_{i=1}^{N}  \left ( \frac{1}{\pi_{i}} -1 \right ) I_{( y_{i} \leq y)}
& = & \frac{\overline{X}_{N}}{f_{N}} \frac{1}{N} \sum_{i=1}^{N} \frac{1}{x_{i}}  I_{( y_{i} \leq y)} - F_N (y) \nonumber \\
\,  & \rightarrow & \left ( \frac{\mathbb{E} [X_1]}{f} K_{-1} (y) -1  \right ) F(y) , \label{eq:limite_b1na} \\
 \frac{F_N (y)^2}{N} \sum_{i=1}^{N}  \left ( \frac{1}{\pi_{i}} -1 \right ) & = & F_N (y)^2 \left (
\frac{\overline{X}_{N}}{f_{N}} \frac{1}{N} \sum_{i=1}^{N} \frac{1}{x_{i}}  -1 \right ) \nonumber \\
\,  & \rightarrow & F(y)^2  \left ( \frac{\mathbb{E} [X_1 ]}{f} \mathbb{E} [ X_1^{-1} ] -1 \right ) ,
\label{eq:limite_b1nb} \\
\frac{F_N (y)}{N} \sum_{i=1}^{N}  \left ( \frac{1}{\pi_{i}} -1 \right ) I_{( y_{i} \leq y)} & \rightarrow &
F(y)^2  \left \{ \frac{\mathbb{E} [X_1 ]}{f} K_{-1} (y) -1 \right \}
\label{eq:limite_b1nc}
\end{eqnarray}
\noindent as $N$ tends to infinity, a.s.-$\mathbb{P}$. From $( \ref{eq:limite_b1na} )$-$( \ref{eq:limite_b1nc} )$, it follows that
\begin{eqnarray}
B_{1,N}(y) & \rightarrow & \left ( \frac{\mathbb{E} [X_1 ]}{f}  K_{-1} (y) -1 \right ) F(y) (1- F(y)) \nonumber \\
\, & \, & -
\frac{\mathbb{E} [X_1 ]}{f} F(y)^2 \left ( K_{-1} (y) - \mathbb{E} [ X_1^{-1} ] \right )
 \;\;
{\mathrm{as}} \; N \rightarrow \infty, \;\; a.s.- \mathbb{P} . \label{eq:limite_b1}
\end{eqnarray}
In the same way, using Lemmas $\ref{lemma_1}$, $\ref{lemma_2}$, it is possible to see that
\begin{eqnarray}
B_{2,N}(y) & = & \frac{\left ( \frac{1}{N} \sum_{i=1}^{N} (1- \pi_i ) (  I_{( y_{i} \leq y )} - F_N (y) )
\right )^2}{\frac{1}{N} \sum_{i=1}^{N} \pi_i (1- \pi_i )} \nonumber \\
\, & \rightarrow & \frac{f^2}{d} \left ( 1- \frac{K_{1} (y)}{\mathbb{E} [X_{1}]} \right )^2  F(y)^2
 \;\;
{\mathrm{as}} \; N \rightarrow \infty, \;\; a.s.- \mathbb{P}  ,
\label{eq:limite_b2} \\
B_{3,N}(y) & = & -2  B_{2,N} \nonumber \\
\, & \rightarrow & -2 \frac{f^2}{d} \left ( 1- \frac{K_{1} (y)}{\mathbb{E} [X_{1}]} \right )^2  F(y)^2
 \;\;
{\mathrm{as}} \; N \rightarrow \infty, \;\; a.s.- \mathbb{P}  .
\label{eq:limite_b3}
\end{eqnarray}
From $( \ref{eq:limite_b1} )$-$( \ref{eq:limite_b3} )$, result $( \ref{eq:limite_s2n} )$ easily follows.
\end{proof}

\begin{lemma}
\label{lemma_4}
For every positive $\epsilon$, with $\mathbb{P}$-probability $1$ there exists an integer $N_\epsilon$ such that
\begin{eqnarray}
\left \vert Z_{i,N} (y) \right \vert \leq \epsilon \pi_i S_N(y) \;\; \forall \, N \geq N_\epsilon .
\label{eq:ineq_lemma4}
\end{eqnarray}
\end{lemma}
\noindent
\begin{proof}[{\bf Proof of Lemma \ref{lemma_4}}]
Let $( (y_i , \, x_i ); \; i \geq 1)$ be a sequence satisfying Lemmas $\ref{lemma_1}$-$\ref{lemma_3}$ (the set of such sequences does have
$\mathbb{P}$-probability $1$), and let $\epsilon >0$ ``small''. Then, there exists $N_\epsilon \geq 1$ (depending on the whole sequence
$( (y_i , \, x_i ); \; i \geq 1)$) such that
\begin{eqnarray}
\frac{S_{N} (y)}{\sqrt{N}} & > &
\left ( \frac{\mathbb{E} [ X_{1} ]}{f} K_{-1} (y) -1 \right ) F(y) (1- F(y) ) -
\frac{\mathbb{E} [ X_{1} ]}{f} \left ( K_{-1} (y) - \mathbb{E} [ X_{1}^{-1} ] \right ) F(y)^2 \nonumber \\
& - &
\frac{f^{2}}{d} \left ( 1- \frac{K_{1} (y)}{\mathbb{E} [ X_{1}]}  \right )^2 F(y)^2 - \epsilon \;\; \forall \, N \geq N_\epsilon ,
\label{eq:ineq4-1} \\
\left \vert Z_{i,N} (y) \right \vert & \leq & 1 + \epsilon +
\frac{f}{\mathbb{E} [ X_{1} ]} X_i \frac{f \left ( 1- K_{1} (y) / \mathbb{E} [ X_1 ] \right)}{d} F(y) \;\; \forall \, N \geq N_\epsilon  .
\label{eq:ineq4-2}
\end{eqnarray}
From $( \ref{eq:ineq4-1} )$ the inequalities
\begin{eqnarray}
\epsilon \pi_i S_N (y) & \geq & \frac{\epsilon}{2} \frac{f}{\mathbb{E} [ X_{1} ]} X_i \left \{
\left ( \frac{\mathbb{E} [ X_{1} ]}{f} K_{-1} (y) -1 \right ) F(y) (1- F(y) )  \right . \nonumber \\
& \, & \left . -
\frac{\mathbb{E} [ X_{1} ]}{f} \left ( K_{-1} (y) - \mathbb{E} [ X_{1}^{-1} ] \right ) F(y)^2 -
\frac{f^{2}}{d} \left ( 1- \frac{K_{1} (y)}{\mathbb{E} [ X_{1}]}  \right )^2 F(y)^2 - \epsilon
\right \} \sqrt{N} \nonumber \\
\, & > & \left (
1 + \epsilon +
\frac{f}{\mathbb{E} [ X_{1} ]} X_i \frac{f \left ( 1- K_{1} (y) / \mathbb{E} [ X_1 ] \right)}{d} F(y) \right )
N^\gamma
\label{eq:ineq4-3}
\end{eqnarray}
\noindent hold, with $0 < \gamma < 1/2$ and for every $N \geq N_\epsilon$. Inequalities
$( \ref{eq:ineq4-2} )$ and $( \ref{eq:ineq4-3} )$ prove
$( \ref{eq:ineq_lemma4} )$.
\end{proof}

\begin{lemma}
\label{lemma_5}
Let $\epsilon$ be a positive number, and let
\begin{eqnarray}
A_N ( \epsilon ) & = & \left \{ i \in \mathcal{U}_N : \; \left \vert Z_{i,N} (y) \right \vert > \epsilon \pi_i S_N(y)
\right \} , \nonumber \\
L_N ( \epsilon )^2 & = & \sum_{i \in A_N ( \epsilon )} \left ( \frac{1}{\pi_{i}} -1  \right ) Z_{i,N} (y)^2 .
\nonumber
\end{eqnarray}
Then
\begin{eqnarray}
\lim_{N \rightarrow \infty} \frac{L_N ( \epsilon )^2}{S_N^2(y)} =0 \;\; a.s.- \mathbb{P} , \; \forall \, \epsilon >0.
\end{eqnarray}
\end{lemma}
\noindent
\begin{proof}[{\bf Proof of Lemma \ref{lemma_5}}]
Immediate consequence of Lemmas $\ref{lemma_3}$, $\ref{lemma_4}$.
\end{proof}

\begin{lemma}
\label{lemma_5b}
Conditionally on $\yN$, $\xN$, as $N$ increases the r.v.
\begin{eqnarray}
\frac{1}{N} \sum_{i=1}^{N} \frac{D_{i}}{\pi_{i}}
\label{eq:rv-weaklaw}
\nonumber
\end{eqnarray}
\noindent tends in probability to $1$,  a.s.-$\mathbb{P}$.
\end{lemma}
\noindent
\begin{proof}[{\bf Proof of Lemma \ref{lemma_5b}}]
The expectation of $( \ref{eq:rv-weaklaw} )$ w.r.t. the sampling design, and conditionally on $\yN$, $\xN$ is
equal to $1$. As far as the variance is concerned, we have first
\begin{eqnarray}
V_P \left (  \left . \frac{1}{N} \sum_{i=1}^{N} \frac{D_{i}}{\pi_{i}} \, \right \vert \YN , \, \XN \right ) & = &
\frac{1}{N^{2}} \left \{ \sum_{i=1}^{N} \frac{1}{\pi_{i}^{2}} V_P ( D_i \, \vert \YN , \, \XN ) \right . \nonumber \\
\, & \, & + \left . \sum_{i=1}^{N} \sum_{j \neq i} \frac{1}{\pi_{i} \pi_{j}} C_P ( D_i , \, D_J \, \vert  \YN , \, \XN )
\right \}  \nonumber \\
\, & \leq & \frac{1}{N^{2}} \left \{ \sum_{i=1}^{N} \frac{1}{\pi_{i}} + \sum_{i=1}^{N} \sum_{j \neq i} \left \vert \frac{\pi_{ij} - \pi_{i} \pi_{j}}{\pi_{i} \pi_{j}} \right \vert \right \} . \nonumber
\end{eqnarray}
As an easy consequence of Lemma $\ref{lemma_2}$, the r.v. $N^{-1} \sum \pi_i^{-1}$ converges a.s.$\mathbb{P}$. Furthermore, the assumption of maximal asymptotic entropy of the sampling design implies (cfr. \cite{hajek81}, Th. 7.4) that
\begin{eqnarray}
\left \vert \frac{\pi_{ij} - \pi_{i} \pi_{j}}{\pi_{i} \pi_{j}} \right \vert \leq \frac{C}{N}
\nonumber
\end{eqnarray}
\noindent $C$ being an absolute constant. This proves the lemma.
\end{proof}

\noindent
\begin{proof}[{\bf Proof of Proposition \ref{asympt_hajek}}]
The proof is based on Lemmas $ \ref{lemma_1}$-$ \ref{lemma_5}$, and it rests on the same ideas as the proof of
Proposition 1 in \cite{conti12}. For this reason, it is only sketched.
First of all, it is not difficult to see that the limiting law of the process  $ ( W_{N}^{H} ( \cdot) ; \; N \geq 1) $
coincides with the limiting law of $ ( \sqrt{f} \widetilde{W}_{N}^{H} ( \cdot) ; \; N \geq 1) $, where
\begin{eqnarray}
\widetilde{W}_{N}^{H} (y) = \frac{1}{\sqrt{N}} \sum_{i=1}^{N} \frac{D_i}{\pi_i} \left ( I_{( y_i \leq y )} - F_N (y) \right ) .
\label{eq:equiv-asy1}
\end{eqnarray}
\noindent
Hence, using Lemmas $ \ref{lemma_1}$-$ \ref{lemma_5}$
and \cite{hajek64} (see also Section 2 of \cite{visek79} and Theorem 1 in \cite{berger98}), it is seen that
the asymptotic distribution of $\widetilde{W}_{N}^{H} (y)$ is normal with mean zero and variance $ f^{-1} C^{H} (y, \, y) $.
 The same kind of result holds for all finite-dimensional distributions of $\widetilde{W}_{N}^{HT} ( \cdot )$, as a consequence of the
 Cram\'{e}r-Wold device.

As far as the tightness is concerned, using the same reasoning as in \cite{conti12} we can confine ourselves to the conditional Poisson sampling design. We have
\begin{eqnarray}
\, & \, & P_{P_{R}} \left ( \left . \vert \widetilde{W}_{N}^{H} (y) - \widetilde{W}_{N}^{H} (s) \vert > \epsilon , \;
\vert \widetilde{W}_{N}^{H} (z) - \widetilde{W}_{N}^{H} (y) \vert > \epsilon \right \vert  \right ) \nonumber \\
\, & \, & \; \leq \frac{1}{N^2 \epsilon^4} V_{P_{R}} \left ( \left . \sum_{i=1}^{N} \frac{D_i - \pi_i}{\pi_i} I_{(s < y_i \leq y)} \right \vert
n_s =n \right ) \, V_{P_{R}} \left ( \left . \sum_{i=1}^{N} \frac{D_i - \pi_i}{\pi_i} I_{(y < y_i \leq z)} \right \vert
n_s =n \right ) . \;\;\;\;\;\; \nonumber \\
\, & \, & \; \leq \left ( \frac{1}{N} \sum_{i=1}^{N} \left ( \frac{1}{\pi_i} -1 \right ) I_{(s < y_i \leq y)}
+ C ( F_N (y) - F_N (s) )^2 \right ) \nonumber \\
\, & \, & \; \times
\left ( \frac{1}{N} \sum_{i=1}^{N} \left ( \frac{1}{\pi_i} -1 \right ) I_{(z < y_i \leq y)}
+ C ( F_N (z) - F_N (y) )^2 \right )
\nonumber \\
\, & \, & \; \leq Q   ( F_N (y) - F_N (s) ) ( F_N (z) - F_N (y) ) ,
\label{eq:tight_A1}
\end{eqnarray}
\noindent with $\mathbb{P}$-probability 1,  $C$, $Q$ being appropriate constants.
Finally, using the Glivenko-Cantelli theorem and the right continuity of $F$, from $( \ref{eq:tight_A1} )$ it follows that
\begin{eqnarray}
\, & \, & P_{P_{R}} \left ( \left . \vert \widetilde{W}_{N}^{H} (y) - \widetilde{W}_{N}^{H} (s) \vert > \epsilon , \;
\vert \widetilde{W}_{N}^{H} (z) - \widetilde{W}_{N}^{H} (y) \vert > \epsilon \right \vert  \right ) \nonumber \\
\, & \, & \; \leq R ( F (y) - F  (s) ) ( F  (z) - F  (y) ) \;\; \forall \, N \geq 1
\label{eq:tight_A2}
\end{eqnarray}
\noindent with $\mathbb{P}$-probability 1, $R$ being an appropriate constant. Inequality $( \ref{eq:tight_A2} )$ proves the
tightness part, and this completes the proof.
\end{proof}

\noindent
\begin{proof}[{\bf Proof of Proposition \ref{lemma_6a}}]
To prove $( \ref{eq:equiv_N*} )$, observe first that
\begin{eqnarray}
E\left [  \frac{N^{*}}{N} \, \vert \DN , \, \yN , \, \XN \right ] & = & K_{1N} \frac{1}{N} \sum_{i=1}^{N} \frac{D_{i}}{\pi_{i}}
 \nonumber \\
\, & \rightarrow & 1
\end{eqnarray}
\noindent as $N$ increases, in probability w.r.t. $\DN$ and for almost all $y_i$s, $x_i$s.
In the second place:
\begin{eqnarray}
V \left ( \left . \frac{N^{*}}{N} \, \right \vert \xN , \, \yN \right )  & \leq &
\frac{1}{N^{2}} \sum_{i=1}^{N} \frac{D_{i}}{\pi_{i}} K_{2N} ( \DN , \, \YN , \, \XN )
\nonumber \\
 \, & + & \frac{c}{N^{3}} K_{3N} ( \DN , \, \YN , \, \XN ) \sum_{i=1}^{N} \sum_{j \neq i } \frac{D_{i}}{\pi_{i}} \frac{D_{j}}{\pi_{j}}
 \nonumber
\end{eqnarray}
From P1-P3 and $( \ref{eq:cond_conv} )$ it is simple to see that
\begin{eqnarray}
V \left ( \left . \frac{N^{*}}{N} \, \right \vert \xN , \, \yN \right ) \rightarrow 0 \;\; as \; N \rightarrow \infty
\nonumber
\end{eqnarray}
in probability w.r.t. $\DN$ and for almost all $y_i$s, $x_i$s, from which $( \ref{eq:equiv_N*} )$ follows.
\end{proof}

\begin{lemma}
\label{lemma_6}
Under assumptions A1-A6, P1-P3,
conditionally on $\yN$, $\xN$, $\DN$, as $N$ increases the statements of Lemmas $\ref{lemma_1}$-$\ref{lemma_5}$ hold true for the predicted population,
and for almost all $y_i$s, $x_i$s values, and in probability w.r.t. $\DN$ and $P_{pred}$.
\end{lemma}
\noindent
\begin{proof}[{\bf Proof of Lemma \ref{lemma_6}}]
Consider the quantity:
\begin{eqnarray}
d^*_{N^{*}} = \sum_{k=1}^{N^{*}} \pi_k^* ( 1- \pi_k^* ) . \label{eq:dNstar_1}
\end{eqnarray}
First of all, using the symbols already introduced, it is not difficult to see that
\begin{eqnarray}
\frac{d^{*}_{N^{*}}}{N} & = & f_N - f_N^2 \frac{\sum_{i=1}^{N} D_{i} \frac{N^{*}_{i}}{N} x_{i}^{2}}{\left ( \sum_{i=1}^{N} D_{i} \frac{N^{*}_{i}}{N} x_{i} \right )^{2}} .
\label{eq:dNstar_2}
\end{eqnarray}
Furthermore, from $( \ref{eq:mean_miltinom} )$-$( \ref{eq:cov_miltinom} )$, it follows that
\begin{eqnarray}
E \left [ \left . \sum_{i=1}^{N} D_{i} \frac{N^{*}_{i}}{N} x_{i} \, \right \vert \DN , \, \YN , \, \XN \right ]  & = &
K_{1N} (\DN , \, \YN , \, \XN ) \sum_{i=1}^{N} D_{i} \pi_i^{-1} x_i
\label{eq:mean_a}
\end{eqnarray}
\noindent and
\begin{eqnarray}
V \left ( \left . \sum_{i=1}^{N} D_{i}   \frac{N^{*}_{i}}{N} x_{i} \, \right \vert \DN , \, \YN , \, \XN \right )
 & = &
\sum_{i=1}^{N} D_{i} x_i^2  \frac{V( N^*_i \, \vert \DN , \, \YN , \, \XN  )}{N^2} \nonumber \\
\,
& + & \sum_{i=1}^{N} \sum_{j \neq i} D_i D_j x_i x_j \frac{C( N^*_i , \, N^*_j \, \vert \DN , \, \YN , \, \XN  )}{N^2} \nonumber \\
\, & \leq & \frac{c}{N} \max ( K_{2N} (\DN , \, \YN , \, \XN ) , K_{3N} (\DN , \, \YN , \, \XN ) ) \nonumber \\
\, & \, & \times \left \{
\frac{1}{N} \sum_{i=1}^{N} D_i x_i + \left (  \frac{1}{N} \sum_{i=1}^{N} D_i x_i \right )^2 \right \} .\label{eq:var_ineq1}
\end{eqnarray}
Using exactly the same arguments as in Propositions $\ref{asympt_hajek}$, $\ref{glivenko_hajek}$, it is now seen that, with $\mathbb{P}$-probability 1 and
in probability w.r.t. the sampling design ({\em i.e.} w.r.t. $\DN$),
\begin{eqnarray}
\frac{1}{N} \sum_{i=1}^{N} D_i x_i & \rightarrow & f \frac{\mathbb{E} [X^2 ]}{\mathbb{E} [ X ]} \nonumber \\
\frac{1}{N} \sum_{i=1}^{N} D_i x_i^{-1} & \rightarrow & \frac{f}{\mathbb{E} [ X ]} \nonumber
\end{eqnarray}
as $N$ increases. From these results it follows that
\begin{eqnarray}
E \left [ \left . \sum_{i=1}^{N} D_{i} \frac{N^{*}_{i}}{N} x_{i} \, \right \vert \DN , \, \YN , \, \XN \right ] & \rightarrow & \mathbb{E} [ X_1 ]
\label{eq:converge1} \\
V \left ( \left . \sum_{i=1}^{N} D_{i}   \frac{N^{*}_{i}}{N} x_{i} \, \right \vert \DN , \, \YN , \, \XN \right )
& \rightarrow & 0 \label{eq:converge2}
\end{eqnarray}
\noindent as $N$ increases, again for almost all $y_i$s, $x_i$s values, and in probability w.r.t. $\DN$.
From $( \ref{eq:converge1} )$, $( \ref{eq:converge2} )$, and Lemma $\ref{lemma_6}$, it is not difficult to conclude that, conditionally on $\DN$, $\yN$, $\xN$,
for almost all $y_i$s, $x_i$s values, and in probability w.r.t. $\DN$, $P_{pred}$,
\begin{eqnarray}
\frac{d^{*}_{N}}{N^{*}} \rightarrow f - f^2 \frac{\mathbb{E} [ X_1^2 ]}{\mathbb{E} [ X_1 ]^2}
\label{eq_converge2bis}
\end{eqnarray}
\noindent that coincides with $( \ref{eq:val_d} )$.

The same arguments can be used to show that Lemmas $\ref{lemma_2}$-$\ref{lemma_5}$ still hold when the actual population is replaced by the predicted population, conditionally on $\yN$, $\xN$, $\DN$, for a set of $y_i$s, $x_i$s having $\mathbb{P}$-probability 1 and for a set of $\DN$s of (design) probability tending to $1$ as $N$, increases,
and where convergence is in probability w.r.t. the (random) mechanism generating the predicted population, {\em $P_{pred}$}. This ends the proof.
\end{proof}

\noindent
\begin{proof}[{\bf Proof of Proposition \ref{prop_resampling_comp}}]
Let
$$
R^{*}_{n} (z) = Pr_{P^{*}} \left ( \left .  Z^{*}_{n,m} \leq z  \, \right
 \vert \yN , \, \xN , \, \DN , \, N^*_1 , \, \dots N^*_M  \right )
$$
be the (resampling) d.f. of
$Z^{*}_{n,m} $ $( \ref{eq:def_z} )$. By Dvoretzky-Kiefer-Wolfowitz inequality (cfr. \cite{massart90}), we have first
\begin{eqnarray}
Pr \left ( \left . \sup_{z} \left \vert \widehat{R}^{*}_{n,M} (z) - R^{*}_{n} (z)  \right \vert  > \epsilon \,
\right \vert \yN , \, \xN , \, \DN , \, N^*_1 , \, \dots N^*_M \right ) \leq 2 \exp \left \{ -2 M \epsilon^2   \right \} .
\label{eq:dkw_ineq}
\end{eqnarray}
\noindent Using the Borel-Cantelli first lemma, and taking into account that $R^{*}_{n} (z)$ converges uniformly to
$\Phi_{0, \sigma^2_\theta} (z)$, $( \ref{eq:risult_1} )$ immediately follows. Statement $( \ref{eq:risult_2} )$
follows from $( \ref{eq:risult_1} )$ and the strict monotonicity of $\Phi_{0, \sigma^2_\theta} (z)$.
Finally, $( \ref{eq:risult_3} )$ is a consequence of Theorem 2.5.5. in \cite{sensinger:93} (pp. 90-91).
\end{proof}

\begin{proof}[{\bf Proof of Proposition \ref{prop_NCPP}}]
First of all, it is immediate to see that
\begin{eqnarray}
\sum_{i=1}^{N} N^{*CPP}_i D_i = N . \nonumber
\end{eqnarray}
\noindent Furthermore, from
\begin{eqnarray}
E [ N^{*CPP}_i \, \vert \DN , \, \YN, \, \XN ] & = & N D_i \pi_i^{-1} / \sum_{j=1}^{N} D_j \pi_j^{-1} \label{eq:mean_cps} \\
V ( N^{*CPP}_i \, \vert \DN , \, \YN, \, \XN ) & = & N \left ( D_i \pi_i^{-1} / \sum_{j=1}^{N} D_j \pi_j^{-1} \right )
\left ( 1- D_i \pi_i^{-1} / \sum_{j=1}^{N} D_j \pi_j^{-1} \right )   \label{eq:var_cps}
\end{eqnarray}
\noindent conditions $P1$ and $P2$ are fulfilled. Finally, if $\tau_{ih}$ are the second order inclusion probabilities of the conditional Poisson sampling, from Th. 7.4 in \cite{hajek81} it follows that
\begin{eqnarray}
\left \vert C ( N^{*CPP}_i , \, N^{*CPP}_h \,  \vert \DN , \, \YN, \, \XN ) \right \vert = \left \vert \tau_{ih} - \tau_i \tau_h \right \vert
\leq \frac{c}{n} \pi_i \pi_h (1+ o(1))
\label{eq:cov_cps}
\end{eqnarray}
\noindent $c$ being an absolute constant, and hence also condition $P3$ is satisfied.
\end{proof}

\begin{proof}[{\bf Proof of Proposition \ref{prop_calib}}]
Take a positive real $\epsilon$, and consider the relaxed optimization problem:
\begin{eqnarray}
\ipotc{{\mathrm{min}}  \;\; \sum_{i \in \s} (N^*_i - N^{*CPP}_i)^2}{N^* =N}{\left \vert
\overline{X}^*_N - \overline{X}_N \right \vert \leq \epsilon}{N^*_i
\geq 0}
\label{eq:probl_minrelax}
\end{eqnarray}
As a consequence of already seen results, the r.v.
\begin{eqnarray}
\sqrt{n} \left ( N^{-1} \sum N^{*CPP}_i D_i x_i - \overline{X}_N \right )
\nonumber
\end{eqnarray}
\noindent tends in distribution to a normal variate with zero mean and positive variance, so that
\begin{eqnarray}
Pr \left ( \left \vert ( \left . N^{-1} \sum N^{*CPP}_i D_i x_i - \overline{X}_N \right \vert \YN , \, \XN , \, \DN  \right \vert > \epsilon
\right )  = O_p ( n^{-1/2} )  \label{eq:conv_ncpp}
\end{eqnarray}
\noindent for each positive $\epsilon$. In other terms, up to a term $O_p ( n^{-1/2} )$ the relaxed problem
$( \ref{eq:probl_minrelax} )$ possesses solution $N^*_i = N^{*CPP}_i$. Since the optimum of the relaxed problem
$( \ref{eq:probl_minrelax} )$ is continuous in $\epsilon$ (cfr. \cite{sundaram96}, Sect. 9.2.2), by letting $\epsilon$ go to $0$
slowly enough as $N$, $n$ increase, it is seen that up to a term $O_p ( n^{-1/2} )$ the solution of the optimization problem
$( \ref{eq:probl_min} )$ is $N^*_i = N^{*CPP}_i$. Result $( \ref{eq:conv_nstar} )$ now follows by observing that
\begin{eqnarray}
 \frac{N^*_i}{N} = \frac{N^{*CPP}_i}{N}  + o_p (1) .
\nonumber
\end{eqnarray}
\end{proof}

\begin{proof}[{\bf Proof of Proposition \ref{prop_hotdeck}}]
As a consequence of Proposition 4 in \cite{marcontsc08}, it is possible to see that the joint d.f. of $X$ and $Y$ in the actual finite population
\begin{eqnarray}
H_N (x, \, y) = \frac{1}{N} \sum_{i=1}^{N} I_{(x_{i} \leq x)} I_{(y_{i} \leq y)}
\nonumber
\end{eqnarray}
\noindent and the joint d.f. of $X$ and $Y$ in the pseudo-population
\begin{eqnarray}
H^{HD}_N (x, \, y) = \frac{1}{N} \sum_{i=1}^{N} I_{(x^{*}_{i} \leq x)} I_{(y^{*}_{i} \leq y)}
\nonumber
\end{eqnarray}
\noindent tend in probability to the same limiting d.f., as $n$, $N$ increase. Hence, the joint d.f. of the
hot-deck pseudo-population, $H^{HD}_N (x, \, y)$, tends to coincide with the actual $H_N (x, \, y)$. This is enough, in its turn,
to conclude that the resampling distribution of $\sqrt{n} ( \theta ( \widehat{F}^{*}_H ) - \theta ( F^{*}_{N^*}  ))$
tends to the same limit as the sampling distribution of $\sqrt{n} ( \theta ( \widehat{F}_H ) - \theta ( F_N  ))$
\end{proof}

\end{document}